\documentclass[english,aip,jap,reprint]{revtex4-1}
\usepackage[T1]{fontenc}
\usepackage[utf8]{inputenc}
\usepackage{xcolor}
\usepackage{pdfcolmk}
\usepackage{babel}
\usepackage{array}
\usepackage{textcomp}
\usepackage{multirow}
\usepackage{amsmath}
\usepackage{amssymb}
\usepackage{graphicx}
\usepackage{wasysym}
\PassOptionsToPackage{normalem}{ulem}
\usepackage{ulem}
\usepackage[unicode=true,pdfusetitle,
 bookmarks=true,bookmarksnumbered=false,bookmarksopen=false,
 breaklinks=false,pdfborder={0 0 0},backref=false,colorlinks=true]
 {hyperref}
\hypersetup{
 allcolors=blue}

\makeatletter

\providecommand{\tabularnewline}{\\}
\providecolor{lyxadded}{rgb}{0,0,1}
\providecolor{lyxdeleted}{rgb}{1,0,0}

\usepackage{textcomp}
\usepackage{siunitx}
\usepackage{dcolumn}
\newcolumntype{g}[1]{D{.}{.}{#1}}
\usepackage{multirow}
\usepackage{graphicx}

\makeatother

\begin{document}

\title{Mössbauer parameters of Fe-related defects in group-IV semiconductors:
first principles calculations }

\author{E. Wright}

\affiliation{Department of Physics and I3N, University of Aveiro, Campus Santiago,
3810-193 Aveiro, Portugal}

\author{J. Coutinho}

\email{jose.coutinho@ua.pt}

\affiliation{Department of Physics and I3N, University of Aveiro, Campus Santiago,
3810-193 Aveiro, Portugal}

\author{S. Öberg}

\affiliation{Department of Engineering Sciences and Mathematics, Luleå University
of Technology, Luleå S-97187, Sweden}

\author{V. J. B. Torres}

\affiliation{Department of Physics and I3N, University of Aveiro, Campus Santiago,
3810-193 Aveiro, Portugal}
\begin{abstract}
We employ a combination of pseudopotential and all-electron density
functional calculations, to relate the structure of defects in supercells
to the isomer shifts and quadrupole splittings observed in Mössbauer
spectroscopy experiments. The methodology is comprehensively reviewed
and applied to the technologically relevant case of iron-related defects
in silicon, and to other group-IV hosts to a lesser degree. Investigated
defects include interstitial and substitutional iron, iron-boron pairs,
iron-vacancy and iron-divacancy. We find that in general, agreement
between the calculations and Mössbauer data is within a 10\% error
bar. Nonetheless, we show that the methodology can be used to make
accurate assignments, including to separate peaks of similar defects
in slightly different environments.
\end{abstract}


\maketitle

\section{Introduction\label{sec:intro}}

The ubiquitous nature of iron in silicon feedstock, in the Si melt
and in as-grown Si crystals, combined with its high diffusivity and
its strong carrier recombination power (particularly in p-type Si),
make Fe one of the most feared contaminants in electronic- and solar-grade
silicon. In fact, the concentration of atomically dispersed Fe in
solar-Si must be kept below a tolerance threshold of only $10^{12}$~cm$^{-3}$
to ensure minority carrier lifetimes $\tau\apprge1\,\mu$s.\cite{ist06}
This is achieved by \emph{storing} Fe impurities away from active
regions in the form of less harmful precipitates with up to several
$\mu$m in diameter, which in turn poses the latent threat of particle
dissolution.\cite{buo05} For further details on the subject of iron
in silicon the reader is directed to the seminal reviews of Istratov,
Hieslmair and Weber.\cite{ist99,ist00}

When dissolved in a Si crystal, Fe atoms occur mostly as interstitial
impurities ($\text{\ensuremath{\mathrm{Fe_{i}}}}$) occupying tetrahedral
interstitial sites. This defect is relatively well understood, and
spectroscopic signals obtained through electron paramagnetic resonance
(EPR), electron-nuclear double resonance,\cite{woo60,sie83,web84}
deep-level transient spectroscopy (DLTS),\cite{fei78,wun82} emission
channeling,\cite{wah05,sil13} and Mössbauer spectroscopy (MS)\cite{gun03a,yos06}
have been reported extensively. Interstitial Fe in Si becomes mobile
above room-temperature. In p-type material it diffuses as a positively
charged ion, being readily trapped by negatively charged acceptors
to form Fe-acceptor pairs.\cite{mac05} Two types of iron-acceptor
pairs have been reported, namely one with trigonal symmetry and another
with orthorhombic symmetry, corresponding to stable and metastable
configurations, respectively.\cite{lud62,geh88,spa98,ist99} Based
on the fully reversible $\mathrm{Fe{}_{i}^{+}}+\mathrm{B_{s}^{-}}\leftrightarrows\mathrm{Fe_{i}B_{s}}$
reaction, Zoth and Bergholz\cite{zot90} proposed a method to quickly
estimate the total concentration of Fe$\mathrm{_{i}}$ in Si samples
with a sensitivity of $10^{11}$~cm$^{-3}$, conferring significant
technological relevance to Fe-acceptor pairs.

While it is consensual that iron impurities occupy interstitial sites
under equilibrium conditions, there is convincing evidence for the
existence of substantial concentrations of substitutional Fe (Fe$_{\mathrm{s}}$)
provided by emission channeling\cite{wah05,sil13} and Mössbauer spectroscopy.\cite{gil90,lan92,wey97,wey99,yos02,yos03}
However, and despite many attempts, powerful techniques like EPR or
DLTS have not detected a signal that can be unambiguously assigned
to Fe$_{\mathrm{s}}$. This is rather puzzling, as theory predicts
Fe$_{\mathrm{s}}$ to be a paramagnetic deep acceptor in the negative
charge state.\cite{est08} Since channeling measurements invariably
involve implantation of radioactive probe ions (followed by thermal
anneals), it seems clear that the presence of Fe$_{\mathrm{s}}$ could
result from the interaction between Fe$_{\mathrm{i}}$ and vacancies.
The same argument can be applied to Mössbauer experiments, where Si
vacancies produced either by high-temperature annealing/quenching
or by the implantation of precursor isotopes (e.g. $^{57}$Mn$^{+}$
), could interact with Fe$_{\mathrm{i}}$, ending up with Fe$_{\mathrm{s}}$
defects. However, Mössbauer experiments where Fe was simply introduced
by vacuum-deposition of an Fe layer on the surface of samples which
were not heat treated, apparently shows a strong Fe$_{\mathrm{s}}$-related
signal at room temperature, presumably without a deliberate introduction
of vacancies.\cite{yos07a}

These observations are rather intriguing, particularly considering
that Neutron Activation Analysis (NAA) indicates that the total amount
of Fe closely matches that measured by EPR,\cite{web80} assigned
to the spin-1 state of neutral Fe$_{\mathrm{i}}$. In these experiments
Fe was introduced by evaporation, followed by in-diffusion at 900ºC-1200ºC
and quenching to minimize Fe-precipitation. Furthermore, the site-independent
enthalpy of formation per iron atom in Si obtained from NAA ($\Delta H=2.94$~eV)
compares well with the same quantity measured by EPR ($\Delta H=2.87$~eV),
which is only sensitive to Fe$_{\mathrm{i}}$. Such agreement provides
compelling evidence that, under thermal equilibrium, the vast majority
of isolated Fe impurities occupy interstitial sites. 

Theoretical modeling, namely electronic structure calculations based
on density functional theory, is a powerful way of studying defects
and their electronics. The (apparent) contradictions described above
call for clarification, particularly in regard to the Mössbauer parameters
of Fe-related defects in Si, as well as the equilibrium concentration
of Fe$_{\mathrm{s}}$ (relative to that of Fe$_{\mathrm{i}}$). Mössbauer
spectroscopy is based on the recoilless emission and absorption of
$\gamma$\nobreakdash-radiation from and by the nuclei of bound atoms.\cite{yos13}
In the absence of a magnetic field at the nucleus, the relevant interactions
are electric monopolar and quadrupolar, which are quantified by the
Isomer Shift (IS or $\delta$) and the Quadrupole Splitting (QS or
$\Delta$) respectively.\cite{gre71,gut12,yos13} In essence, these
two parameters are sensitive to the magnitude of the contact electron
density and to the non-sphericity of the density surrounding the Mössbauer
nucleus ($^{57}$Fe in this case).

Motivated by the impact of Fe and its complexes on Si based technology,
\textsuperscript{57}Fe MS has been used to study the location, diffusivity
and electronic activity of elemental Fe impurities.\cite{lan92,ist99,kob00,wey00}
A summary of experimental and calculated IS values for Fe$_{\mathrm{i}}$
and Fe$_{\mathrm{s}}$ defects in Si and other group IV semiconductors
is presented in Table~\ref{tab:ms-survey}. Consistent agreement
between theoretical and experimental values of $\delta$ has enabled
the identification of Fe$_{\mathrm{i}}$ and Fe$_{\mathrm{s}}$ defects
in Si, where $\mathrm{\delta(Fe_{\mathrm{s}})\simeq-}0.04$~mm/s
and $\mathrm{\delta(Fe_{\mathrm{i}})\simeq}0.8$~mm/s.\cite{gun02b,gun03a,yos06}
In Ge, the picture also appears to be well established, with experiments
and theory suggesting $\mathrm{\delta(Fe_{\mathrm{i}})\simeq}0.8$~mm/s
and a small, albeit positive $\mathrm{\delta(Fe_{\mathrm{s}})\simeq}0.06$~mm/s.\cite{gun03b}
In diamond, and despite good agreement between the calculated value
and the $\delta=0.22$~mm/s resonance assigned to Fe$_{\mathrm{i}}$,
calculations for Fe$_{\mathrm{s}}$ severely underestimate the experimental
result by a factor ranging between 5 and 10. Finally, in SiC there
are two possible sites for Fe$_{\mathrm{i}}$ and Fe$_{\mathrm{s}}$
defects. The interstitial impurity can have Si or C first neighbors
( Fe$_{\mathrm{i,Si}}$ or Fe$_{\mathrm{i,C}}$ respectively), while
substitutional Fe can replace Si or C atoms (Fe$_{\mathrm{Si}}$ or
Fe$_{\mathrm{C}}$ respectively). While the agreement between theory
and experiments is reasonable, only a single resonance has been reported
for substitutional Fe, although two peaks are theoretically predicted.
Interestingly, the overall trend of the MS peaks in Si, Ge, Diamond
and SiC,\cite{gun03b,wey04b,bha08} is characterized by a linear increase
of $\delta(\mathrm{Fe_{s})}$ and $\mathrm{\delta(Fe_{i})}$ with
the distance between Fe and its first neighbors, as depicted in Figure~2
of Ref.~\onlinecite{gun06}.

\begin{table}
\caption{\label{tab:ms-survey}Summary of experimental (Exp.) and calculated
(Calc.) isomer shifts for Fe$_{\mathrm{i}}$ and Fe$_{\mathrm{s}}$
defects in group IV semiconductors, in mm/s. For SiC, distinct rows
are used for the values of Fe with Si (Fe$_{\mathrm{i,Si}}$/Fe$_{\mathrm{C}}$)
and C (Fe$_{\mathrm{i,C}}$/Fe$_{\mathrm{Si}}$) first neighbors.
NA stands for \emph{not available}. }

\begin{tabular}{ccccccccccc}
\hline 
 &  & \multicolumn{2}{c}{Fe$_{\mathrm{i}}$} &  &  &  &  &  & \multicolumn{2}{c}{Fe$_{\mathrm{s}}$}\tabularnewline
 &  & Exp. & Calc. &  &  &  &  &  & Exp. & Calc.\tabularnewline
\hline 
C &  & 0.22\footnotemark[1]  & 0.22\footnotemark[2] &  &  &  &  &  & $-$0.91\footnotemark[1] & $-$0.19-0.09\footnotemark[2]\tabularnewline
 &  &  &  &  &  &  &  &  &  & \tabularnewline
\multirow{2}{*}{SiC} & (Fe$_{\mathrm{i,Si}}$/Fe$_{\mathrm{C}}$) & 0.67\footnotemark[3]  & 0.49\footnotemark[4] &  &  &  &  &  & NA & $-$0.55\footnotemark[4]\tabularnewline
 & (Fe$_{\mathrm{i,C}}$/Fe$_{\mathrm{Si}}$) & 0.33\footnotemark[3] & 0.27\footnotemark[4]  &  &  &  &  &  & $-$0.23\footnotemark[3] & $-$0.24\footnotemark[4]\tabularnewline
 &  &  &  &  &  &  &  &  &  & \tabularnewline
\multirow{2}{*}{Si} &  & 0.76/0.77\footnotemark[5] & 0.72\footnotemark[6] &  &  &  &  &  & $-$0.04\footnotemark[5] & $-$0.06\footnotemark[6]\tabularnewline
 &  & 0.808\footnotemark[7] & 0.89\footnotemark[8] &  &  &  &  &  & $-$0.043\footnotemark[7] & 0.13\footnotemark[8]\tabularnewline
 &  &  &  &  &  &  &  &  &  & \tabularnewline
Ge &  & 0.80\footnotemark[9]  & 0.78\footnotemark[8] &  &  &  &  &  & 0.059\footnotemark[9] & 0.08\footnotemark[8]\tabularnewline
\hline 
\end{tabular}

\footnotetext[1]{Ref.~\onlinecite{wey04b}}
\footnotetext[2]{Ref.~\onlinecite{bha98}}
\footnotetext[3]{Ref.~\onlinecite{bha08}}
\footnotetext[4]{Corresponding to 6H-SiC Ref.~\onlinecite{elz14}}
\footnotetext[5]{Varies with doping, cf. Ref.~\onlinecite{gun03a}}
\footnotetext[6]{Ref.~\onlinecite{abr14}}
\footnotetext[7]{Ref.~\onlinecite{yos06}}
\footnotetext[8]{Ref.~\onlinecite{kub93}}
\footnotetext[9]{Ref.~\onlinecite{gun03b}}
\end{table}

The electronic activity of $\mathrm{Fe_{s}}$ and $\mathrm{Fe_{i}}$
defects in Si has been studied using MS by varying the type and concentration
of dopants in the samples.\cite{gun02a,gun03a,yos07a} However, conflicting
MS parameters have been reported for some elemental Fe defects in
Si. The ISOLDE consortium reported a $\mathrm{\delta(Fe_{\mathrm{s}})}=-0.04$~mm/s
peak which is effectively independent of the dopant type and concentration,
and was therefore assigned to the neutral charge state of substitutional
iron (Fe$\mathrm{_{s}^{0}}$). From the same experiments, neutral
and positively charged $\mathrm{Fe_{i}}$ defects were assigned to
resonances at $\mathrm{\delta(Fe_{i}^{0})}=0.72$~mm/s and $\mathrm{\delta(Fe_{i}^{+})=0.78}$~mm/s
in n-type and p-type material, respectively.\cite{gun02a,gun03a}
More recently, and for the same defects, Yoshida and his group reported
IS values of $\mathrm{\delta(Fe_{\mathrm{s}}^{0})\simeq-}0.17$~mm/s,
$\mathrm{\delta(Fe_{i}^{0})}=0.40$~mm/s and $\mathrm{\delta(Fe_{i}^{+})=0.80}$~mm/s.\cite{yos07a,yos16} 

Other defects were tentatively assigned using MS, namely a quadrupole-split
doublet labeled Fe$_{\mathrm{N}}$ with $\Delta(\mathrm{Fe_{N})}=0.51$~mm/s
and $\mathrm{\delta(Fe_{\mathrm{N}})=}0.43$~mm/s, assigned to an
iron-vacancy (Fe$_{\mathrm{i}}$V) pair,\cite{gun03a} and another
doublet labeled $\mathrm{Fe_{\mathrm{D}}}$ which has been associated
to Fe in regions damaged by the ion-implantation process, with $\Delta(\mathrm{Fe_{D})}=1.02$~mm/s
and $\mathrm{\delta(Fe_{\mathrm{D}})=}0.33$~mm/s.\cite{gun02b} 

\emph{Ab-initio} calculations of Mössbauer parameters have played
an important role in the interpretation and validation of experimental
data. For the particular case of defects in semiconductors, these
calculations have been hampered by the fact that solid state effects
can only be accounted for if defects are embedded in a sufficiently
large supercell or cluster. Owing to the computational effort involved,
few attempts to calculate the IS and QS of defects were made,\cite{kub93,bha98}
which invariably implied several limitations and approximations like
the use of small supercells, a local or semi-local approach to the
electronic exchange-correlation interactions, or a non-relativistic
treatment of core states. However, state-state-of-the-art calculations
employing (linearized) augmented plane-wave methods (eventually complemented
by local orbitals), can account for relativistic core and valence
states, as well as anisotropy effects in the core potentials. Encouraging
results have been recently reported, including the calculation of
$\mathrm{\delta(Fe_{s})}$ and $\mathrm{\delta(Fe_{i})}$ in SiC,\cite{elz14}
and for more elaborate Fe-related defects in Si such as Fe-vacancy
and Fe-interstitial complexes.\cite{abr14} The aim of this work is
to review the methodology used to calculate Mössbauer parameters from
first-principles, and to apply the protocol to a set of relevant Fe
defects in Si. In Section~\ref{sec:theoretical-method} we describe
the methodology used to calculate formation energies, electronic levels,
contact densities and electric field gradients. We then proceed to
calculate the Mössbauer calibration constants (Section~\ref{sec:calibration}).
In Sections~\ref{sec:iron-defects-in-si} and \ref{sec:other-group-iv}
we report the results obtained for Fe defects in Si and other group-IV
semiconductors, namely formation energies, electronic levels and Mössbauer
parameters, followed by our conclusions in Section~\ref{sec:conclusions}.

\section{Theoretical method\label{sec:theoretical-method}}

All calculations were carried out within density functional theory,\cite{hoh64,koh65}
adopting the generalized gradient approximation to the exchange-correlation
potential among electrons as proposed by Perdew, Burke and Ernzerhof
(PBE).\cite{per96,per97} Mössbauer spectroscopy probes the shape
of the electron density at nuclear sites, depending on both core and
valence states. This means that the calculation of the Mössbauer parameters
requires an accurate description of the coupling between core and
valence states, and therefore involves solving an all-electron problem.
To this end, contact densities $\bar{n}$ (averaged within the nuclear
volume) and electric field gradients $V_{ij}$ (EFG) were calculated
using the full-potential all-electron $\mathtt{ELK}$ code, employing
a basis of augmented plane-waves plus local orbitals (APW+lo).\cite{elk15}
On the other hand, for the purpose of describing the geometry and
chemistry of Fe defects, core electrons can be safely \emph{frozen}
and considered as if they were bound to free atoms. In this case,
ground state structures of supercells with and without defects, their
respective energies and defect levels were obtained using the planewave
$\mathtt{VASP}$ code\cite{kre93,kre94,kre96a,kre96b} with core states
treated within the projector augmented-wave (PAW) method.\cite{blo94,kre99}
Supercell structures obtained within the PAW method were subsequently
plugged into the all-electron code, in order to extract $\bar{n}$
and $V_{ij}$ values for the Fe defects.

The band structures were sampled over $\Gamma$-centered $N_{1}\!\times\!N_{2}\!\times\!N_{3}$
grids of $\mathbf{k}$-points along the $\mathbf{b}_{1}$, $\mathbf{b}_{2}$
and $\mathbf{b}_{3}$ reciprocal lattice vectors, or MP-$N_{1}\!\times\!N_{2}\!\times\!N_{3}$
in abbreviated form.\cite{mon76} The grids employed for each case
are reported throughout the text, alongside a description of the corresponding
system.

\subsection{PAW calculations\label{sub:paw}}

For these calculations, PAW potentials (constructed with a specific
valence configuration) for $\mathrm{Fe}(3s^{2}3p^{6}3d^{7}4s^{1})$,
$\mathrm{B}(2s^{2}2p^{1})$, $\mathrm{C}(2s^{2}2p^{2})$, $\mathrm{Si}(3s^{2}3p^{2})$
and $\mathrm{Ge}(4s^{2}4p^{2})$ were considered. Kohn-Sham valence
states were expanded in plane-waves with kinetic energies up to $E_{\mathrm{cut}}=450$~eV.
The electronic spin treatment was collinear and allowed to relax.
Atomic coordinates were relaxed using either a conjugate-gradient
method or a quasi-Newton algorithm, until the maximum force acting
on atoms was not larger than $2.5\times10^{-3}$~eV/Å.

Iron-related point defects were inserted into 64- and 216-atom supercells
with a simple cubic lattice. Unless otherwise specified, isomer shifts
and quadrupole splittings were calculated for Fe defects in the smaller
supercells, while other defect properties (formation energies, electrical
levels, transformation barriers) were calculated using the larger
supercells. The lattice parameters were those that minimized the energy
of bulk primitive cells, namely $a_{\mathrm{C}}=3.5736$~Å, $a_{\mathrm{Si}}=5.4687$~Å,
$a_{\mathrm{Ge}}=5.7829$~Å and $a_{\mathrm{SiC}}=4.3785$~Å, for
diamond, Si, Ge and 3C-SiC (cubic polytype), respectively. As expected
from the use of the PBE functional, these figures overestimate (by
less than 1\%, except for Ge where the deviation is about 2\%) the
corresponding experimental lattice parameters of 3.5668~Å, 5.4310~Å,
5.6579~Å and 4.3596~Å.\cite{mad96} For the case of Fe defects in
Si-rich SiGe alloys, the host consisted of a Si supercell with one
substitutional Ge atom. The lattice parameter $a_{\mathrm{SiGe}}=5.4736$~Å
was scaled linearly between $a_{\mathrm{Si}}$ and $a_{\mathrm{Ge}}$
assuming a Vegard alloying regime. The band structure of all defective
supercells was integrated by using MP-$2^{3}$\ $\mathbf{k}$-point
grids, totaling 8 reducible points. For all group-IV semiconductor
primitive cells we employed MP-$12^{3}$\ $\mathbf{k}$-point grids.

\begin{figure}
\includegraphics[width=8cm]{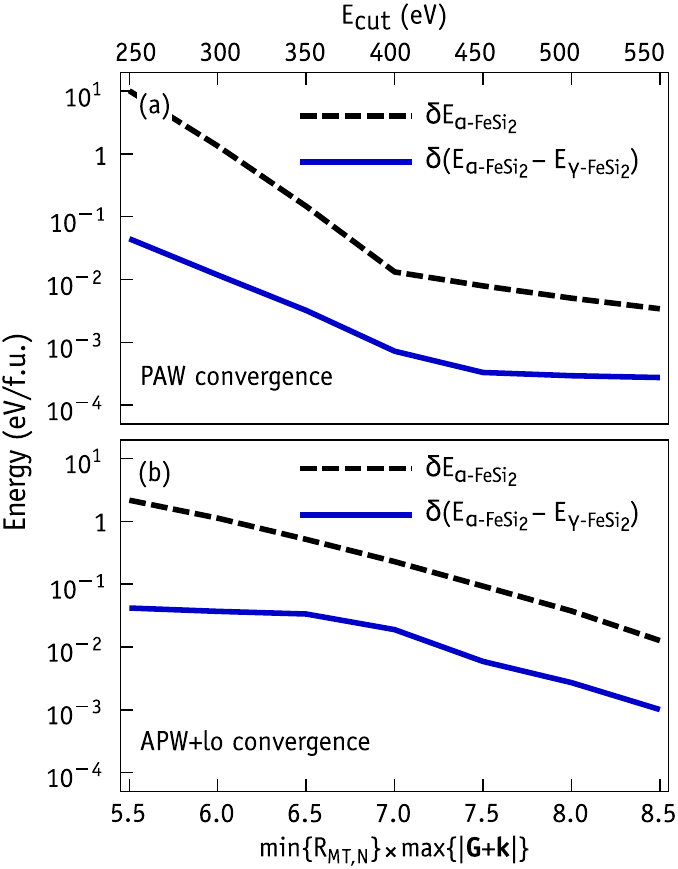}

\caption{\label{fig:test-ecut}Convergence of the total energy of $\alpha$-FeSi$_{2}$
(dashed line) against the convergence of the energy difference between
$\alpha$-FeSi$_{2}$ and $\gamma$-FeSi$_{2}$ crystals (solid line)
as a function of the basis size. (a) PAW and (b) APW+lo calculations.
Energies are in eV per formula unit (eV/f.u.). The values were calculated
with respect to well converged results with plane wave cut-offs for
the valence wavefunctions set at $E_{\mathrm{cut}}=600$~eV for the
PAW calculations and $E_{\mathrm{cut}}=250$~eV (corresponding to
$\min\left\{ R_{\mathrm{MT},N}\right\} \times\max\left\{ \left|\mathbf{G}+\mathbf{k}\right|\right\} =9.0$)
in the APW+lo calculations. See text for further details and definitions.}
\end{figure}

Total energies and energy differences were tested as a function of
$E_{\mathrm{cut}}$ and $\mathbf{k}$-point sampling in order to ensure
that the latter converged within 1~meV. For the sake of example,
the dashed line depicted in Figure~\ref{fig:test-ecut}(a) represents
the variation of the total energy per formula unit (in eV/f.u.) of
$\alpha$-FeSi$_{2}$ as a function of the plane-wave cut-off energy.
Specifically, it represents $\delta E_{\alpha\text{-}\mathrm{FeSi_{2}}}=|E_{\alpha\text{-}\mathrm{FeSi_{2}}}(E_{\mathrm{cut}})-E_{\alpha\text{-}\mathrm{FeSi_{2}}}(600\,\mathrm{eV})|$,
where $E_{\alpha\text{-}\mathrm{FeSi_{2}}}(E_{\mathrm{cut}})$ is
the energy of crystalline $\alpha$-FeSi$_{2}$ as a function of $E_{\mathrm{cut}}$.
Also in the same figure, the solid line shows the variation of the
energy difference (in eV/f.u.) between the $\alpha$-FeSi$_{2}$ and
$\gamma$-FeSi$_{2}$ phases as a function of $E_{\mathrm{cut}}$,
with respect to analogous calculations with $E_{\mathrm{cut}}=600$~eV.
While the total energy converges within 10~meV around $E_{\mathrm{cut}}\apprge400$~eV,
the energy difference is at least one order of magnitude more accurate
for the same value of $E_{\mathrm{cut}}$. The Brillouin zone (BZ)
integration for the $\alpha$ and $\gamma$ phases of FeSi$_{2}$
was carried out on MP-$16^{2}\times14$ and MP-$16^{3}$ grids of\ \textbf{k}-points,
respectively.

\subsection{APW+lo calculations\label{sub:apw}}

The contact densities and EFG tensors for Fe nuclei were calculated
using the APW+lo method.\cite{mad01,sin06} We addressed the all-electron
problem self-consistently within density functional theory. Core states
were treated relativistically by solving the spherical Dirac equation,
while valence states were expanded in APW functions subject to the
full potential. The APWs are dual-representation functions that divide
the cell volume into two regions, namely a \emph{muffin-tin} (MT)
region comprising non-overlapping spheres with radius $R_{\mathrm{MT},N}$
centered at each nuclear coordinate $\mathbf{R}_{N}$, and an interstitial
(I) region which is basically elsewhere in the volume $\Omega$ of
the periodic cell. Electronic states $\psi_{\mathbf{k}\lambda}^{\sigma}(\mathbf{r})$,
where $\sigma$ and $\lambda$ denote spin and band index, are then
expanded using a basis of augmented plane waves,

\begin{equation}
\phi_{\mathbf{G}+\mathbf{k}}(\mathbf{r})=\begin{cases}
{\displaystyle \sum_{Njlm}}a_{\mathbf{G}+\mathbf{k}}^{Njlm}\,u_{Nl}^{(j)}(r_{N},\epsilon_{Njl})\,Y_{lm}(\hat{\mathbf{r}}_{N}), & r_{N}<R_{\mathrm{MT},N}\\
\Omega^{-1/2}\exp\left[\,\mathrm{i}\left(\mathbf{G}+\mathbf{k}\right)\cdot\mathbf{r}\right], & \mathbf{r}\in\mathrm{I}.
\end{cases}\label{eq:apw-basis}
\end{equation}
Each partial wave function enclosed by the MT sphere results from
the product of a spherical harmonic $Y_{lm}$ and the $j$-th energy
derivative of the radial solution to the Schrödinger equation for
the free atom $N$, $u_{Nl}^{(j)}=\partial^{j}u_{Nl}/\partial\epsilon^{j}$,
with $l$ and $m$ being the usual angular and magnetic moment quantum
numbers. In this work we considered contributions of APWs with $l\leq8$.
Each radial function's derivative $u_{Nl}^{(j)}$ depends parametrically
on a linearization energy $\epsilon_{Njl}$, and is a function of
a local coordinate $\mathbf{r}_{N}=r_{N}\,\hat{\mathbf{r}}_{N}=\mathbf{r}-\mathbf{R}_{N}$
with origin at the $N$-th nucleus. The $a_{\mathbf{G}+\mathbf{k}}^{Njlm}$
coefficients are chosen such that each partial wave function matches
a plane wave counterpart at the MT/I boundary, where $\mathbf{G}$
and $\mathbf{k}$ are a reciprocal lattice vector and a wave vector
within the first BZ (special $\mathbf{k}$-point), respectively. Note
that according to this definition, the Linearized APW (LAPW) scheme\cite{and75}
is readily available by considering only $u_{Nl}$ and its first energy
derivative $\partial u_{Nl}/\partial\epsilon$ in the partial wave
expansion. For each calculation the basis was specified by the condition
$\min\left\{ R_{\mathrm{MT},N}\right\} \times\max\left\{ \left|\mathbf{G}+\mathbf{k}\right|\right\} =8$,
and the muffin-tin radii for Fe, C, Si, Ge, F, Br and Ti set to 1.11~Å,
0.95~Å, 1.11~Å, 1.27~Å, 0.95~Å, 1.27~Å, and 1.27~Å, respectively.

Further basis flexibility was conferred to the APWs by adding a set
of $lm$-dependent local orbitals centered on each nucleus,

\begin{equation}
\phi_{\mathrm{lo}}^{Nlm}(\mathbf{r}_{N})=\sum_{j}a_{\mathrm{lo}}^{Njlm}\,u_{Nl}^{(j)}(r_{N},\epsilon_{Njl})\,Y_{lm}(\hat{\mathbf{r}}_{N}),
\end{equation}
where the coefficients $a_{\mathrm{lo}}$ are determined by normalization
and requiring that $\phi_{\mathrm{lo}}$ functions (and eventually
their radial derivatives) vanish at the MT boundary. All basis functions
and linearization energies were those provided by the official \texttt{ELK}
distribution. The potential (and the electron density) also employed
a dual representation. Within the MT region is was expanded using
lattice harmonics with angular momentum $l_{\phi}\leq7$, while across
the interstitial region it was expanded in stars of planewaves whose
kinetic energies were limited at $E_{\mathrm{cut,\phi}}=1.96$~keV.

Looking at Figure~\ref{fig:test-ecut}(b) it is clear that for a
particular basis specification, and analogously to the PAW calculations,
energy differences are about one order of magnitude better converged
than the total energies. The figure also shows that under \emph{production}
conditions ($\min\left\{ R_{\mathrm{MT},N}\right\} \times\max\left\{ \left|\mathbf{G}+\mathbf{k}\right|\right\} =8$)
the energy differences converge within a few meV. The $\gamma$-phase
of FeSi$_{2}$ is metastable with respect to the $\alpha$-phase by
$E_{\text{\ensuremath{\gamma}-FeSi2}}-E_{\text{\ensuremath{\alpha}-FeSi2}}=0.37$~eV/f.u.
These figures are in agreement with $E_{\text{\ensuremath{\gamma}-FeSi2}}-E_{\text{\ensuremath{\alpha}-FeSi2}}=0.32$~eV/f.u.
reported from previous all-electron LAPW calculations.\cite{mor99}
Other convergence tests showed that total energies and energy differences
varied by less than 2~meV and 0.3~meV with respect to more demanding
conditions in describing the potential using $l_{\phi}=9$ and $E_{\mathrm{cut,\phi}}=2.67$~keV.

\subsection{Calculation of isomer shifts and quadrupole splittings}

For any point $\mathbf{r}$ far from the nucleus, the potential $\phi(\mathbf{r})$
due to a nuclear charge density $\rho_{\mathrm{N}}(\mathbf{r})$ can
be represented by a multipolar expansion $\phi\equiv\phi_{0}+\phi_{i}+\ldots+\phi_{N}$.
Each of the resulting $\mbox{\ensuremath{\phi}}_{i}$ terms interacts
with the electronic charge density, adding $E_{i}$ energy terms to
the total energy.\cite{gre71} Since the electric monopole term $\phi_{0}$
does not account for the finite nature of the nucleus, this method
can not describe the electronic contact density and its contribution
to $E_{0}$. Relativistic and perturbative methods have been employed
to solve this problem, where $\rho_{\mathrm{N}}$ was described by
an equivalent uniformly charged sphere of radius $R_{\mathrm{N}}$.\cite{bod53}
Accordingly, by treating the corresponding potential as a perturbation
to $\mbox{\ensuremath{\phi}}_{0}$, a first order energy correction

\begin{equation}
\Delta E_{0}=\frac{2\pi}{5}\,Z\,R_{\mathrm{N}}^{2}\,\bar{n}
\end{equation}
is introduced, where $Z$ is the atomic number, and $\bar{n}$ is
the electron contact density due to $s$-electrons. Assuming that
the change in the nuclear radius upon a $\gamma$ transition is $\Delta R_{\mathrm{N}}\ll R_{\mathrm{N}}$,
the corresponding shift in the transition energy is\cite{bod53,shi64}

\begin{equation}
\Delta E_{\gamma}\simeq\frac{4\pi}{5}\,Z\,R_{\mathrm{N}}^{2}\,\frac{\Delta R_{\mathrm{N}}}{R_{\mathrm{N}}}\,\bar{n}.
\end{equation}
Now, if we consider that $\rho_{\mathrm{N}}$ is unaltered by the
surrounding electron density, \emph{i.e.} that $R_{\mathrm{N}}$ and
$\Delta R_{\mathrm{N}}$ depend on $Z$ only, $\Delta E_{\gamma}$
becomes purely dependent on $\bar{n}$, \emph{i.e.} on the chemical
environment of the nucleus. The isomer shift $\delta$ is defined
as the relative difference between values of $\Delta E_{\gamma}$
for two distinct chemical environments, such that\cite{ang04}

\begin{eqnarray}
\delta & = & \frac{c}{E_{\gamma}}\,\frac{4\pi}{5}\,Z\,R_{\mathrm{N}}^{2}\,\frac{\Delta R_{\mathrm{N}}}{R_{\mathrm{N}}}\left(\bar{n}_{\mathrm{a}}-\bar{n}_{\mathrm{s}}\right)\\
\nonumber \\
 & = & \alpha\,\left(\bar{n}_{\mathrm{a}}-\bar{n}_{\mathrm{s}}\right),\label{eq:isomershift}
\end{eqnarray}
where $\bar{n}_{\mathrm{a}}-\bar{n}_{\mathrm{s}}$ is the relative
contact density between the \emph{absorber} and \emph{source} isomers
used in velocity-scanning experiments, and $\alpha$ is a proportionality
factor known as the IS calibration constant. As $\delta$ is a relative
quantity, a reference material must be chosen for which $\delta=0$~$\mathrm{m\cdot s^{-1}}$.
For the 14.4~keV transition in $\mbox{\ensuremath{\mathrm{^{57}Fe}}}$
that is usually $\alpha$\nobreakdash-Fe (a ferromagnet with body
centered cubic structure), and hence $\bar{n}_{\mathrm{s}}\equiv\bar{n}(\text{\ensuremath{\alpha}-Fe})$.
Thus, provided that $\alpha$ is known, isomer shifts may be readily
estimated from calculated values of $\bar{n}_{\mathrm{a}}$ and $\bar{n}_{\mathrm{s}}$.
Details of the calculation of $\alpha$ are reported in Section~\ref{sec:calibration}.

\begin{figure}
\includegraphics{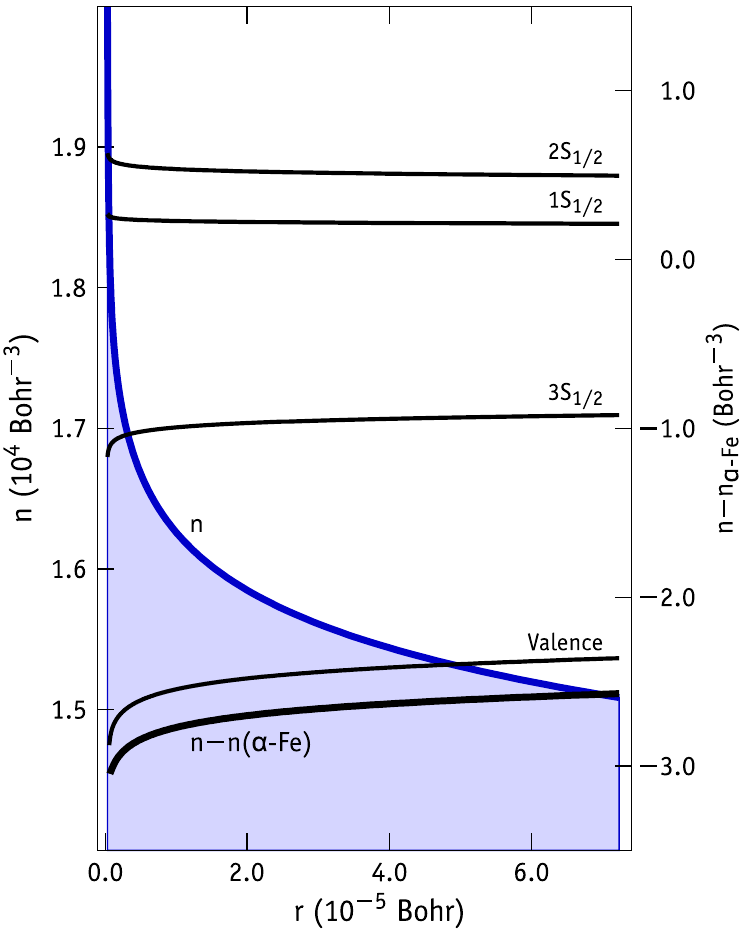}

\caption{\label{fig:rho-core-method}(a) Electron density, $n$, of a neutral
Fe$_{\mathrm{i}}$ impurity in a 64-Si atom supercell (blue curve)
as a function of the radial distance from the center of the $^{57}$Fe
nucleus (with calculated radius $R_{\mathrm{N}}=7.1214\times10^{-5}\ \text{Bohr}$).
Also shown is the relative density of the neutral Fe$_{\mathrm{i}}$
with respect to the $\alpha$-Fe source, $n-n_{\text{\ensuremath{\alpha}-Fe}}$.
Contributions to the relative density from $NS_{1/2}$ core states
($N=1$, 2 and 3) and valence states are represented as thinner curves.}
\end{figure}

The calculation of the contact density involves integrating the electron
density within a sphere of radius $R_{\mathrm{N}}$, here assumed
to be,\cite{ang04}
\begin{equation}
R_{\mathrm{N}}=\left(R_{0}+\dfrac{R_{1}}{A^{2/3}}+\dfrac{R_{2}}{A^{4/3}}\right)A^{1/3},
\end{equation}
with $R_{0}=0.9071$~fm, $R_{1}=1.105$~fm and $R_{2}=-0.548$~fm,
and $A$ is the atomic mass number. For $^{57}$Fe this gives $R_{\mathrm{N}}=3.7685$~fm
($7.1214\times10^{-5}$~Bohr). For the sake of example, let us briefly
describe the calculation of the relative contact density in Equation~\ref{eq:isomershift}.
Figure~\ref{fig:rho-core-method} depicts the radial electron density
centered on a neutral Fe impurity located at a tetrahedral interstitial
site in a 64-Si atom supercell. The horizontal axis spans the integration
limits $3.9223\times10^{-7}\,\text{Bohr}\leq r\leq7.1214\times10^{-5}\,\text{Bohr}$
used to evaluate all contact densities reported in this paper. The
thick curve at the bottom represents the relative density on the Fe$_{\mathrm{i}}$
impurity with respect to $\alpha$-Fe. These correspond to contact
densities $\bar{n}=15985.189$~Bohr$^{-3}$ and $\bar{n}(\text{\ensuremath{\alpha}-Fe})=15987.914$~Bohr$^{-3}$,
indicating a more diffuse density on the Fe impurity in Si with $\bar{n}-\bar{n}(\text{\ensuremath{\alpha}-Fe})=-2.725$~Bohr$^{-3}$.
It is interesting to note that by changing the environment of an Fe
atom from metallic $\alpha$-Fe to a tetrahedral interstitial site
in Si, most changes in the contact density come from the high-energy
states, namely $3S_{1/2}$ and valence.

It is noted that experimental IS values include a second order Doppler
contribution as $\delta_{\mathrm{Exp}}=\delta+\delta_{\mathrm{D}}$,
which within the harmonic approximation takes the form of $\delta_{\mathrm{D}}=-E/2mc$,
where $E$ is the time-averaged kinetic energy of the resonant nucleus
with oscillating mass $m$ and $c$ is the speed of light in the vacuum.
At high temperatures, from the equipartition principle, $E\approx3k_{\mathrm{B}}T/2$
for both absorber and source ($k_{\mathrm{B}}$ being the Boltzmann
constant), and to first order they mutually cancel in Eq.~\ref{eq:isomershift}.
On the other hand, at low temperatures $E$ is approximately the zero-point
energy of the resonant atom. We may estimate the impact of neglecting
the Doppler contribution to relative IS shifts of Fe defects in Si
by looking at the highest vibrational frequencies in the phonon density
of states. For the source ($\alpha$-Fe), the highest allowed phonon
frequency is about $\nu=9.3$~THz ($\hbar\omega\approx40$~meV).\cite{mau14}
Although zero-point motion energies for Fe defects in Si are not available,
we make use of the highest phonon frequency in $\beta$-FeSi$_{2}$,
\emph{i.e.}, $\nu=15$~THz ($\hbar\omega\approx62$~meV),\cite{tan10}
which is the most stable phase for iron disilicide. Accordingly, for
a vibrating $^{57}$Fe nucleus in both materials, the difference between
Doppler contributions is only about 0.03~mm/s.

For a state with nuclear spin $I>1/2$ the nuclear charge density
is aspherical and the expansion of $\phi(\mathbf{r})$ includes a
quadrupolar term $\phi_{2}$, usually denoted $Q$, which interacts
electrostatically with the local EFG.\cite{gre71} The EFG is a traceless
tensor, $V_{xx}+V_{yy}+V_{zz}=0$, so, by adopting the convention
whereby $|V_{xx}|\leq|V_{yy}|\leq|V_{zz}|$, the asymmetry parameter
\begin{equation}
\eta=(V_{xx}-V_{yy})/V_{zz},
\end{equation}
and $V_{zz}$ are sufficient to describe $V$ in the principal axis
system. As $I=3/2$ for the excited state in the $14.4$~keV transition
in $\mbox{\ensuremath{\mathrm{^{57}Fe}}}$, the quadrupolar interaction
results in a doublet with separation given by the quadrupole splitting\cite{gre71}
\begin{equation}
\Delta=\frac{Qc}{2E_{\gamma}}V_{zz}\sqrt{1+\frac{\eta^{2}}{3}},\label{eq:quadsplit}
\end{equation}
where $V_{zz}\sqrt{1+\eta^{2}/3}$ is often termed the \emph{effective
electric field gradient}, $V_{\mathrm{eff}}$. The sign of the experimental
quadrupole splitting cannot always be determined, although it is possible
to do so in some cases, \emph{e.g.} by further splitting the doublet
via a magnetic interaction or by considering the relative intensities
of the lines as a function of the orientation of the crystalline sample.\cite{reu02}
Equation~\ref{eq:quadsplit} allows us to estimate quadrupole splittings
from calculated $V_{\mathrm{eff}}$ values, provided that we know
the nuclear quadrupole moment $Q$ for the $^{57}$Fe nucleus. Although
this has been previously calculated as $Q=0.17$~b (Barn units) from
first-principles,\emph{\cite{duf95,wdo07}} we reproduce the calibration
procedure in Section~\ref{sec:calibration} below.

\subsection{Formation energy of Fe defects in SiC\label{sub:formation-energy}}

Formation energies of defects were all calculated within the PAW method
according to the usual approach,\cite{nor93}

\begin{equation}
E_{\mathrm{f}}=E_{\mathrm{def}}-\sum_{i}n_{i}\mu_{i},
\end{equation}
where charge neutrality is assumed, and $E_{\mathrm{def}}$ is the
total energy of a supercell with $n_{i}$ atoms of species $i$ whose
chemical potential $\mu_{i}\equiv\partial G_{\mathrm{std}}/\partial N_{i}$.
Here $G_{\mathrm{std}}$ is the Gibbs free energy of an ensemble of
atoms in some \emph{standard} phase containing $N_{i}$ elements.
Calculating $\mu_{i}$ is quite straightforward for a defect in a
homopolar semiconductor. For instance, in the case of an iron impurity
in Si, $\mu_{\mathrm{Si}}$ is readily calculated from the energy
per atom in a pristine Si unit cell, while $\mu_{\mathrm{Fe}}$ can
be estimated from the energy per Fe atom in $\beta$-FeSi$_{2}$ as
$\mu_{\mathrm{Fe}}=(E_{\text{\ensuremath{\beta}-FeS\ensuremath{i_{2}}}}-16\mu_{\mathrm{Si}})/8$
where $E_{\text{\ensuremath{\beta}-FeS\ensuremath{i_{2}}}}$ is the
energy per primitive cell of the disilicide comprising 8 FeSi$_{2}$
formula units. The result will effectively give us the formation energy
of the Fe impurity in thermodynamic equilibrium with a source/sink
of Fe in the form of $\beta$-FeSi$_{2}$.

For a compound semiconductor like SiC the above formalism becomes
more elaborate as $\mu_{\mathrm{Si}}$ and $\mu_{\mathrm{C}}$ can
vary within certain limits. For instance, assuming that the standard
phases of Si and C are crystalline silicon and diamond with respective
chemical potentials $\mu_{\mathrm{Si}}^{0}$ and $\mu_{\mathrm{C}}^{0}$,
we have $\mu_{\mathrm{Si}}\leq\mu_{\mathrm{Si}}^{0}$ and $\mu_{\mathrm{C}}\leq\mu_{\mathrm{C}}^{0}$,
otherwise the SiC crystal would not be stable and would separate into
its elemental phases. On the other hand we have

\begin{equation}
\mu_{\mathrm{Si}}+\mu_{\mathrm{C}}=\mu_{\mathrm{SiC}}=\mu_{\mathrm{Si}}^{0}+\mu_{\mathrm{C}}^{0}-H_{\mathrm{f,SiC}},\label{eq:chempot1}
\end{equation}
where $\mu_{\mathrm{SiC}}$ is the energy per formula unit in the
SiC crystal and $H_{\mathrm{f,SiC}}$ is the heat of formation of
SiC. Hence, we can arrive at

\begin{eqnarray}
\mu_{\mathrm{Si}} & = & \left(\mu_{\mathrm{SiC}}+\Delta\mu\right)/2\\
\mu_{\mathrm{C}} & = & \left(\mu_{\mathrm{SiC}}-\Delta\mu\right)/2,
\end{eqnarray}
with $\Delta\mu_{\mathrm{SiC}}^{0}-H_{\mathrm{f,SiC}}\leq\Delta\mu\leq\Delta\mu_{\mathrm{SiC}}^{0}+H_{\mathrm{f,SiC}}$
and $\Delta\mu_{\mathrm{SiC}}^{0}=\mu_{\mathrm{Si}}^{0}-\mu_{\mathrm{C}}^{0}$,
where the lower and upper limits of $\Delta\mu$ account for a SiC
crystal grown under Si-rich conditions ($\Delta\mu=\Delta\mu_{\mathrm{SiC}}^{0}-H_{\mathrm{f,SiC}}$)
or C-rich conditions ($\Delta\mu=\Delta\mu_{\mathrm{SiC}}^{0}+H_{\mathrm{f,SiC}}$).
Finally, to obtain $H_{\mathrm{f,SiC}}$ we simply use Eq.~\ref{eq:chempot1}
along with values for $\mu_{\mathrm{Si}}^{0}$, $\mu_{\mathrm{C}}^{0}$
and $\mu_{\mathrm{Si\mathrm{C}}}$ calculated from bulk cells with
their respective equilibrium lattice parameters. This gives $H_{\mathrm{f,SiC}}=0.54$~eV,
which compares fairly well with the experimental value of $0.68$~eV.\cite{crc14}

\subsection{Electrical Levels}

Electrical levels were calculated using the marker method.\cite{res99}
This approach is based on a comparison between electron affinities
($A$) or ionization energies ($I$) of a particular defect under
scrutiny, with analogous $A$'s and $I$'s calculated for a reference
system (referred to as \emph{marker}). Accordingly, a donor level
with respect to the valence band top is then obtained as

\begin{eqnarray}
E_{\mathrm{def}}(q/q+1)-E_{\mathrm{v}} & = & I_{\mathrm{def}}(q/q+1)-I_{\mathrm{mark}}(q/q+1)+\nonumber \\
 & + & \left\{ E_{\mathrm{mark}}(q/q+1)-E_{\mathrm{v}}\right\} _{\mathrm{exp}},
\end{eqnarray}
where the $I$ values are calculated from total energies $E(q)$ for
specific charge states, $I(q/q+1)=E(q)-E(q+1)$, for the examined
defect ($I_{\mathrm{def}}$) and marker ($I_{\mathrm{mark}}$), and
finally the term between curly brackets is an experimental quantity
and refers to the donor level of the marker with respect to $E_{\mathrm{v}}$.
Analogously, for acceptor levels we have,

\begin{eqnarray}
E_{\mathrm{c}}-E_{\mathrm{def}}(q-1/q) & = & A_{\mathrm{mark}}(q-1/q)-A_{\mathrm{def}}(q-1/q)-\nonumber \\
 & - & \left\{ E_{\mathrm{c}}-E_{\mathrm{mark}}(q-1/q)\right\} _{\mathrm{exp}},
\end{eqnarray}
where electron affinities are calculated as $A(q-1/q)=E(q-1)-E(q).$

Provided that there is a similarity in the shape and localization
between donor (or acceptor) states of the examined defect and marker,
this method is claimed to mitigate well known difficulties in the
calculation of defect levels through error cancelation. The errors
essentially derive from the non-exact treatment of the electronic
exchange-correlation, and from finite-size effects due to insufficiently
large supercell, like electronic dispersion, elastic strain or long-ranged
electrostatic interactions across a lattice of charged defects.\cite{gos07}
In the calculation of the levels of Fe related defects in Si, we adopted
the interstitial Ti impurity as a marker defect. Like most defects
scrutinized below, Ti is an interstitial metallic impurity, and has
levels measured at $E_{\mathrm{c}}-E(-/0)=0.09$~eV, $E(0/+)-E_{\mathrm{v}}=0.87$~eV
and $E(0/+)-E_{\mathrm{v}}=0.26$~eV.\cite{mat89} A handicap of
the marker method is precisely its dependence on a particular measurement.
Since the assignment of the Ti levels was recently questioned,\cite{kol12,sch15}
we double checked the results using interstitial and substitutional
copper marker levels, namely $\mathrm{Cu_{i}}(0/+)$, $\mathrm{Cu_{s}}(-/0)$
and $\mathrm{Cu_{s}}(=/-)$ at $E_{\mathrm{c}}-0.15$~eV, $E_{\mathrm{v}}+0.41$~eV
and $E_{\mathrm{c}}-0.17$~eV, respectively.\cite{bro85,ist97,yar11}Mössbauer
calibration constants\label{sec:calibration}

The IS calibration constant ($\alpha$) and the nuclear quadrupole
moment ($Q$) for the 14.4~keV transition of $^{57}$Fe were obtained
by fitting experimental values of the IS ($\delta_{\text{Exp}}$)
and the QS ($\Delta_{\text{Exp}}$ ) to calculated values of $\bar{n}-\bar{n}(\text{\ensuremath{\alpha}-Fe})$
and $V_{\mathrm{eff}}$, through the respective linear relationships
expressed in Eqs.~\ref{eq:isomershift} and \ref{eq:quadsplit}.
To this end, we chose a collection of Fe-related compounds that cover
a wide range of $\delta_{\text{Exp}}$ and $\Delta_{\text{Exp}}$
values. Since we are interested in the calculation of Mössbauer parameters
of Fe defects in Si, several iron silicides were included in that
collection. We found that considering crystalline structures with
their respective relaxed (theoretical) lattice constants, resulted
in considerable scattering in the linear plots. This is perhaps due
to the rather distinct bonding character among the materials considered,
which leads to non-systematic errors from the exchange-correlation
treatment. Hence, in line with Refs.~\onlinecite{duf95} and \onlinecite{wdo07},
crystalline structures with experimental lattice constants were used
for the calculation of $\alpha$ and $Q$.

\begin{table*}
\caption{\label{tab:calibration}Crystal name, space-group, unit cell lattice
parameters (in Å), lattice site of Fe (when applicable), calculated
relative contact densities ($\bar{n}-\bar{n}(\text{\ensuremath{\alpha}-Fe})$),
experimental isomer shifts ($\delta_{\text{Exp}}$), calculated effective
electric field gradients ($V_{\text{eff}}$), and experimental quadrupole
splittings ($\Delta_{\text{Exp}}$) for Fe nuclei in various compounds
used to find the Mössbauer calibration constants. The sign of the
$\Delta_{\text{Exp}}$ values is shown for the cases where direct
measurements are available. Fe$_{3}$Si and $\beta$-FeSi$_{2}$ crystals
contain two inequivalent Fe atoms ($\mathrm{Fe_{A,C}}$/$\mathrm{Fe_{B}}$
and $\mathrm{Fe_{I}}$/$\mathrm{Fe_{II}}$, respectively).}

\begin{tabular}{llllllr@{\extracolsep{0pt}.}lr@{\extracolsep{0pt}.}lr@{\extracolsep{0pt}.}lr@{\extracolsep{0pt}.}l}
\hline 
 &  &  &  &  &  & \multicolumn{2}{c}{$\bar{n}-\bar{n}(\text{\ensuremath{\alpha}-Fe})$} & \multicolumn{2}{c}{$\delta_{\text{Exp}}$} & \multicolumn{2}{c}{$V_{\text{eff}}$} & \multicolumn{2}{c}{$\Delta_{\text{Exp}}$}\tabularnewline
Crystal & Space Group & Unit cell &  &  & Site & \multicolumn{2}{c}{($\text{Bohr}^{-3}$)} & \multicolumn{2}{c}{(mm/s)} & \multicolumn{2}{c}{($\times10^{21}\,\text{V/\ensuremath{m^{2}}}$)} & \multicolumn{2}{c}{(mm/s)}\tabularnewline
\hline 
$\alpha$-Fe & $Im\bar{3}m$ & $a=2.8601$\cite{ace94,mor09} &  &  &  & \multicolumn{2}{c}{$0.0$} & \multicolumn{2}{c}{$0.0$} & \multicolumn{2}{c}{} & \multicolumn{2}{c}{}\tabularnewline
TiFe & $Pm\bar{3}m$ & $a=2.9789$\cite{tho89} &  &  &  & \multicolumn{2}{c}{$0.6754$} & \multicolumn{2}{c}{$-0.145$\cite{mie75}} & \multicolumn{2}{c}{} & \multicolumn{2}{c}{}\tabularnewline
Fe$_{3}$Si & $Fm\bar{3}m$ & $a=5.653$\cite{nic76} &  &  & $\mathrm{Fe_{A,C}}$ & \multicolumn{2}{c}{$-0.9746$} & \multicolumn{2}{c}{$+0.26$\cite{ham11}} & \multicolumn{2}{c}{} & \multicolumn{2}{c}{}\tabularnewline
 &  &  &  &  & $\mathrm{Fe_{B}}$ & \multicolumn{2}{c}{$-0.2183$} & \multicolumn{2}{c}{$+0.08$\cite{ham11}} & \multicolumn{2}{c}{} & \multicolumn{2}{c}{}\tabularnewline
$\alpha$-FeSi$_{2}$ & $P4/mmm$ & $a=2.6955$\cite{mii15} &  &  &  & \multicolumn{2}{c}{$-0.8494$} & \multicolumn{2}{c}{$+0.202$\cite{reu01}} & \multicolumn{2}{c}{$-4.0398$} & \multicolumn{2}{c}{$-0.730$\cite{reu01}}\tabularnewline
 &  & $c=5.1444$\cite{mii15} &  &  &  & \multicolumn{2}{c}{} & \multicolumn{2}{c}{} & \multicolumn{2}{c}{} & \multicolumn{2}{c}{}\tabularnewline
$\beta$-FeSi$_{2}$ & $Cmca$ & $a=7.791$\cite{dus71} &  &  & $\mathrm{Fe_{I}}$ & \multicolumn{2}{c}{$-0.2972$} & \multicolumn{2}{c}{$+0.076$\cite{fan95}} & \multicolumn{2}{c}{$+2.9768$} & \multicolumn{2}{c}{$+0.525$\cite{fan95}}\tabularnewline
 &  & $b=7.883$\cite{dus71} &  &  & $\mathrm{Fe_{II}}$ & \multicolumn{2}{c}{$-0.3642$} & \multicolumn{2}{c}{$+0.091$\cite{fan95}} & \multicolumn{2}{c}{$-1.6932$} & \multicolumn{2}{c}{$-0.315$\cite{fan95}}\tabularnewline
 &  & $c=9.863$\cite{dus71} &  &  &  & \multicolumn{2}{c}{} & \multicolumn{2}{c}{} & \multicolumn{2}{c}{} & \multicolumn{2}{c}{}\tabularnewline
$\epsilon$-FeSi & $P2_{1}3$ & $a=4.489$\cite{pau48} &  &  &  & \multicolumn{2}{c}{$-1.0545$} & \multicolumn{2}{c}{$+0.282$\cite{fan96b}} & \multicolumn{2}{c}{$+2.7874$} & \multicolumn{2}{c}{$+0.495$\cite{fan96b}}\tabularnewline
FeF$_{2}$ & $P4_{2}/mnm$ & $a=4.6966$ \cite{bal66} &  &  &  & \multicolumn{2}{c}{$-5.44238$} & \multicolumn{2}{c}{$+1.467$\cite{wer67}} & \multicolumn{2}{c}{$+15.166$} & \multicolumn{2}{c}{$+2.85$\cite{wer67}}\tabularnewline
 &  & $c=3.3091$ \cite{bal66} &  &  &  & \multicolumn{2}{c}{} & \multicolumn{2}{c}{} & \multicolumn{2}{c}{} & \multicolumn{2}{c}{}\tabularnewline
FeF$_{3}$ & $R\bar{3}c$ & $a=5.362$\cite{leb85} &  &  &  & \multicolumn{2}{c}{$-1.8808$} & \multicolumn{2}{c}{$+0.489$\cite{wer68}} & \multicolumn{2}{c}{$+0.2486$} & \multicolumn{2}{c}{$0.044$\cite{wer68}}\tabularnewline
 &  & $\alpha=57.94$\cite{leb85} &  &  &  & \multicolumn{2}{c}{} & \multicolumn{2}{c}{} & \multicolumn{2}{c}{} & \multicolumn{2}{c}{}\tabularnewline
FeBr$_{2}$ & $P\bar{3}m1$ & $a=3.772$\cite{wil59} &  &  &  & \multicolumn{2}{c}{$-4.3077$} & \multicolumn{2}{c}{$+1.120$\cite{pfl68}} & \multicolumn{2}{c}{$+6.3969$} & \multicolumn{2}{c}{$1.132$\cite{pfl68}}\tabularnewline
 &  & $c=6.223$\cite{wil59} &  &  &  & \multicolumn{2}{c}{} & \multicolumn{2}{c}{} & \multicolumn{2}{c}{} & \multicolumn{2}{c}{}\tabularnewline
\hline 
\end{tabular}
\end{table*}

Table~\ref{tab:calibration} lists all Fe-compounds employed in the
calibration procedure. $\alpha$-Fe is the ferromagnetic ground state
of iron with body-centered cubic structure ($Im\bar{3}m$ space group).
In line with most Mössbauer experiments, this is considered the reference
substance in the calculation of IS values. The APW+lo calculation
employed a MP-$18^{3}$ grid of special $\mathbf{k}$-points to sample
the BZ. The calculated magnetic moment per primitive cell (per Fe
atom) was $M=2.18\,\mu_{\mathrm{B}}$, where $\mu_{\textrm{B}}$ is
the Bohr magneton. This compares well with the experimental value
$M=2.22\,\mu_{\mathrm{B}}$.\cite{kit05}

TiFe is a metallic compound that crystalizes in the CsCl prototypical
cubic structure ($Pm\bar{3}m$ space group). Although it has two transition
metals per primitive cell, it is a diamagnetic compound. The BZ of
this crystal was sampled with a MP-18$^{3}$ grid of $\mathbf{k}$-points
and the resulting relative contact density was $\bar{n}-\bar{n}(\text{\ensuremath{\alpha}-Fe})=0.6754$~Bohr$^{3}$.
This is consistent with some electron transfer from Ti to Fe as already
reported in Ref.~\onlinecite{pap75}.

The inter-metallic Fe$_{3}$Si solid crystallizes in the $D0_{3}$
structure ($F\bar{m}3m$ space group) and is a ferromagnetic Heusler
compound that has attracted much interest.\cite{ion05} This structure
can be viewed as two inter-penetrating zincblende lattices offsetted
along the cube edge by $a/2$ (where \textbf{$a$} the lattice constant).\cite{nic76}
While one of the sub-lattices comprises two inequivalent Fe atoms
($\mathrm{Fe_{A}}$ and $\mathrm{Fe_{B}}$), the second sub-lattice
is made of a $\mathrm{Fe_{C}}$ and Si atom pair. Importantly, $\mathrm{Fe_{A}}$
and $\mathrm{Fe_{C}}$ are equivalent by symmetry and $\mathrm{Fe_{B}}$
sits on a site that is similar to that in $\alpha$-Fe --- it has
8 equivalent $\mathrm{Fe}$ first neighbors at the corners of a cube.
The BZ was sampled using a MP-16$^{3}$ grid of $\mathbf{k}$-points.
Calculated magnetic moments are $M(\mathrm{Fe_{A/C}})=1.38$~$\mu_{\mathrm{B}}$
and $M(\mathrm{Fe_{B}})=2.58$~$\mu_{\mathrm{B}}$. The latter is
close to that in $\alpha$-Fe and both compare well the experimental
values of 1.35~$\mu_{\mathrm{B}}$ and 2.2-2.4~$\mu_{\mathrm{B}}$,
respectively.\cite{nic83}

$\alpha$-FeSi$_{2}$ is a high-temperature stable ($967^{\circ}\mathrm{C}\lesssim T\lesssim1223\mathrm{^{\circ}C}$)
iron disilicide with tetragonal crystal structure. It is a metallic
and diamagnetic compound that is metastable with respect to the structure
observed at low-temperatures, namely $\beta$-FeSi$_{2}$. The energy
difference between fully relaxed $\alpha$ and $\beta$ phases was
calculated at 0.18~eV/f.u., comparing well with 0.19~eV/f.u. from
previous calculations.\cite{mor99} The $\beta$-FeSi$_{2}$ phase
attracted much attention due to its semiconducting nature with a band
gap of $\sim0.85$~eV and envisaged applications in optoelectronics
and photovoltaics.\cite{leo97} Despite many experimental and theoretical
efforts, there is still no agreement about the characteristics and
the nature of the band gap in this material.\cite{fil96,mor99} $\beta$-FeSi$_{2}$
is a base-centered orthorhombic crystal (space group $Cmca$) with
8 Fe atoms (occupying two inequivalent sites, namely Fe$_{\mathrm{I}}$
and Fe$_{\mathrm{II}}$) and 16 Si atoms per primitive cell.\cite{dus71}
Each Fe species is surrounded by a Jahn-Teller distorted cube of Si
atoms, and the material is non-magnetic. Among the silicides, the
$\beta$-FeSi$_{2}$ is the only one where a non-axially symmetric
EFG tensor is obtained ($\eta>0$), namely, $\eta_{\mathrm{I}}=0.62$
and $\eta_{\mathrm{II}}=0.75$, for Fe$_{\mathrm{I}}$ and Fe$_{\mathrm{II}}$,
respectively. Due to several effects, the $\eta$ parameters could
not be measured reliably. However, they are in qualitative agreement
with previous linear-muffin-tin orbital (LMTO) calculations ($\eta_{\mathrm{I}}=0.36$
and $\eta_{\mathrm{I}}=0.41$).\cite{fan97} Finally, within the Fe-Si
phase diagram, we also considered a monosilicide, namely $\epsilon$-FeSi,
which is also a non-magnetic semi-metallic material. It crystallizes
with a cubic lattice, and contains 4 Fe (Si) atoms per primitive cell
at trigonal sites (space group $P2_{1}3$).\cite{wat63,pau48,fan96b}
For the silicides, BZ sampling grids were MP-$16^{2}\!\times\!14$,
MP-6$^{3}$ and MP-12$^{3}$ for the $\alpha$, $\beta$ and $\epsilon$
phases, respectively.

Three iron halides, respectively FeF$_{2}$, FeBr$_{2}$ and FeF$_{3}$,
were also considered in the collection of calibration compounds. They
are all anti-ferromagnetic insulators and show a wide spectrum of
isomer shifts and quadrupole splittings (in decreasing order as they
are referred). FeF$_{2}$ crystallizes in the rutile structure ($P4_{2}/mnm$
space group), the conventional cell has two formula units where the
Fe$^{2+}$ magnetic moments align anti-parallel along the $c$-axis.
Our calculations indicate that within the muffin-tin sphere of Fe,
the magnetic moments are $M(\mathrm{Fe})=\pm3.79\,\mu_{\mathrm{B}}$,
matching previous PAW calculations,\cite{lop12} and not far from
the experimental figure $M(\mathrm{Fe})=\pm3.93\,\mu_{\mathrm{B}}$.\cite{str04}
FeBr$_{2}$ crystallizes with the CdI$_{2}$ structure ($P\bar{3}m1$
space group), comprising hexagonal layers of Fe atoms sandwiched between
bromine atom layers.\cite{wil59} The magnetic structure of FeBr$_{2}$
comprises alternate anti-parallel magnetized layers of Fe atoms, with
a large measured magnetization $M(\mathrm{Fe})=\pm4.4\,\mu_{\mathrm{B}}$
per Fe$^{2+}$ ion.\cite{wil59} Our calculations account only for
$M(\mathrm{Fe})=3.76\,\mu_{\mathrm{B}}$, perhaps resulting from insufficiencies
in the semi-local treatment of exchange-correlation interactions in
accounting for the van der Waals type bonding between the anti-ferromagnetic
FeBr$_{2}$ layers. Finally, FeF$_{3}$ is a rhombohedral crystal
($R\bar{3}c$ space group) and a canted anti-ferromagnet where Fe$^{3+}$
ions are located at the center of octahedra of six fluorine atoms
which are slightly tilted with respect to the crystallographic axes.\cite{wol58,leb85}
Among the iron halides, only FeF$_{2}$ has a non-axial EFG with a
calculated asymmetry parameter $\eta=0.32$, in reasonable agreement
with the experimental estimate $\eta=0.4$.\cite{wer67} For the iron
halides, BZ sampling grids were MP-$12^{2}\!\times\!16$, MP-$12^{2}\!\times\!4$
and MP-12$^{3}$ for FeF$_{2}$, FeBr$_{2}$ and FeF$_{3}$, respectively.

\begin{figure}
\includegraphics{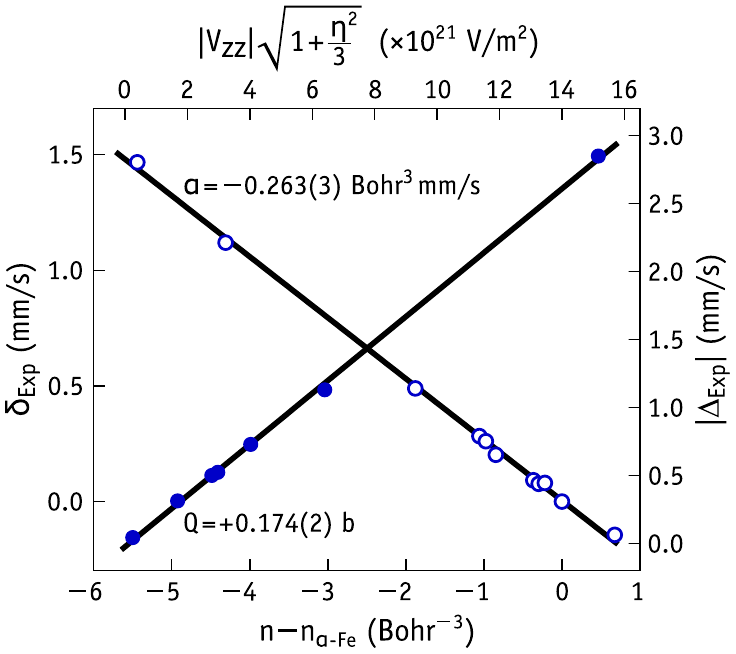}

\caption{\label{fig:calibration}Fit of the isomer shift calibration constant
($\alpha$) and $^{57}$Fe nuclear quadrupole moment ($Q$) using
data from Table~\ref{tab:calibration}. Open symbols represent experimental
isomer shifts ($\delta_{\mathrm{Exp}}$) against the calculated relative
contact densities, $\bar{n}-\bar{n}(\text{\ensuremath{\alpha}-Fe})$.
Closed symbols represent absolute experimental quadrupole splittings,
$|\Delta_{\text{Exp}}|$, plotted against the absolute effective electric
field gradients, $|V_{zz}|(1+\eta^{2}/3)^{1/2}$. Errors quoted within
parentheses derive from the least squares linear regression. Total
errors in $\alpha$ and $Q$ are expected to be 5-10\%.}
\end{figure}

In Figure~\ref{fig:calibration} we show two plots based on the data
from Table~\ref{tab:calibration}, representing experimental isomer
shifts versus calculated relative contact densities (open symbols)
and experimental quadrupole splittings versus calculated effective
EFG's (closed symbols) for the materials described above. Least squares
linear fits to the data resulted in an IS calibration constant $\alpha=-0.26$~Bohr$^{3}$\,mm/s
and in a nuclear quadrupole moment $Q=+0.17$~b. The positive sign
of the nuclear quadrupole moment is deduced from the sign of $\Delta_{\mathrm{exp}}$
for the silicides and FeF$_{2}$. The calculated values for $\alpha$
and $Q$ are in good agreement with previous full-potential APW calculations,
namely those by Wdowik and Ruebenbauer\cite{wdo07} where $\alpha=-0.29$~Bohr$^{3}$mm/s
and $Q=+0.17$~b and those by Dufek \emph{et~al.\cite{duf95}} where
$Q=+0.16$~b was obtained.

\section{Iron defects in silicon\label{sec:iron-defects-in-si}}

\subsection{Solubility and lattice location of dissolved iron in silicon}

In iron-plated samples annealed at high temperatures, the solubility
of in-diffused Fe in the Si bulk depends on the formation of an equilibrium
silicide/Si interface underneath the surface.\cite{web83} Below the
Si-Fe eutectic ($T\lesssim1480$~K), the maximum concentration of
all iron dissolved in the bulk, $[\mathrm{Fe_{Si}}]$, and that in
the silicide phase, $[\mathrm{Fe_{FeSi}}]$, are related as follows 

\begin{equation}
\ln\left([\mathrm{Fe_{Si}}]/[\mathrm{Fe_{FeSi}}]\right)=\Delta S/k_{\mathrm{B}}-\Delta H/k_{\mathrm{B}}T,\label{eq:equilibrium-condition}
\end{equation}
where $\Delta S$ and $\Delta H$ are the excess relative partial
entropy and enthalpy of formation, respectively, resulting from the
transferral of an Fe atom from the silicide to bulk Si. Assuming that
$[\mathrm{Fe_{FeSi}}]$ is temperature independent, for elemental
impurities like tetrahedral interstitial and substitutional iron,
Eq.~\ref{eq:equilibrium-condition} becomes an Arrhenius relation,

\begin{equation}
[\mathrm{Fe_{Si}}]=c_{0}\,\exp\left(\Delta S/k_{\mathrm{B}}-\Delta H/k_{\mathrm{B}}T\right),\label{eq:arrhenius}
\end{equation}
where $c_{0}$ is the number of possible defect sites and orientations
per unit volume in the crystal (for tetrahedral interstitial and substitutional
impurities in Si we have $c_{0}=5\times10^{22}$~cm$^{-3}$). Using
measured values of entropy ($\Delta S/k_{\mathrm{B}}=8.2$) and enthalpy
of formation for dissolved Fe in Si,\cite{web83} one obtains a solubility
$[\mathrm{Fe}]\approx4.7\times10^{15}$~cm$^{-3}$ at $T=1100^{\circ}$C.
While the change in entropy is a quantity that is difficult to estimate,
we can easily calculate the formation enthalpy of Fe$_{\mathrm{i}}$
and Fe$_{\mathrm{s}}$ defects by using appropriate chemical potentials
for Fe and Si species. Assuming that the source of Fe is the most
stable phase of iron disilicide, namely $\beta$-FeSi$_{2}$ (see
Sec.~\ref{sub:formation-energy}), we have $\Delta H(\mathrm{Fe_{i}})=E_{\mathrm{def}}(\mathrm{Fe_{i}})-216\mu_{\mathrm{Si}}-\mu_{\mathrm{Fe}}=2.73$~eV
and $\Delta H(\mathrm{Fe_{s}})=E_{\mathrm{def}}(\mathrm{Fe_{s}})-215\mu_{\mathrm{Si}}-\mu_{\mathrm{Fe}}=3.23$~eV.
These calculations were carried out using 216-Si atom supercells with
respective defects and a $\beta$-FeSi$_{2}$ primitive cell (using
MP-2$^{3}$ and MP-6$^{3}$ special $\mathbf{k}$-point sets, respectively).
We note that the close agreement between $\Delta H(\mathrm{Fe_{i}})$
and the experimental figure from EPR data ($\Delta H=2.87$~eV) suggests
that the iron source in high-temperature in-diffused samples is actually
a $\beta$-FeSi$_{2}$ layer. In fact, considering $\mu_{\mathrm{Fe}}$
from $\alpha$-FeSi$_{2}$ or Fe$_{3}$Si phases, the agreement between
calculated and experimental formation enthalpies worsens considerably,
with $\Delta H(\mathrm{Fe_{i}})=2.55$~eV and 1.80~eV, respectively.

Assuming that formation entropy values of Fe$_{\mathrm{i}}$ and Fe$_{\mathrm{s}}$
(with respect to Fe in $\beta$-FeSi$_{2}$) are dominated by configurational
contributions,\cite{est08} they should be comparable. Hence, from
$[\mathrm{Fe_{s}}]/[\mathrm{Fe_{i}}]=\exp\left[\left(-\Delta H(\mathrm{Fe_{s}})+\Delta H(\mathrm{Fe_{i}})\right)/k_{\mathrm{B}}T\right]$,
and for $T\sim900{}^{\circ}$C-1200$^{\circ}$C we obtain a concentration
of Fe$_{\mathrm{s}}$ which is about 1\%-2\% of the total dissolved
Fe. In Ref.~\onlinecite{web80} the authors reported that within
the above temperature interval, the concentration of Fe$_{\mathrm{i}}$
(detected by EPR) was slightly but invariably lower than the total
Fe in the samples (from NAA). Since some of the iron could have precipitated
during the quenching stage, it was suggested that this imbalance represented
an upper limit for $[\mathrm{Fe_{s}}]$.

It is also interesting to note that Gilles and his co-workers\cite{gil90}
reported the enhancement of the Fe solubility in n$^{+}$-type Si
by up to four orders of magnitude in the temperature range 700$^{\circ}$C-850$^{\circ}$C.
This was attributed to the formation of immobile substitutional Fe
and possibly to pairing with phosphorous. Also from emission channeling
in n-type Si, the concentration of substitutional Fe along with that
of a defect referred to as near bond-centered iron, was found to be
larger than that of Fe$_{\mathrm{i}}$, while the opposite was observed
in p$^{+}$-doped samples, \emph{i.e.} Fe$_{\mathrm{i}}$ was the
most abundant impurity.\cite{sil13} The small difference between
the formation enthalpies of Fe$_{\mathrm{i}}$ and Fe$_{\mathrm{s}}$,
combined with previous predictions that Fe$_{\mathrm{s}}$ is a deep
acceptor,\cite{est08} could explain the measurements referred above.
Nevertheless, there is a clear need for further experimental and theoretical
efforts in order to identify Fe$_{\mathrm{s}}$ and its electronic
levels .

\subsection{Interstitial iron in silicon}

From PAW calculations, we arrived at the following ground states for
Fe interstitial in silicon, $\mathrm{^{1}Fe_{i}^{0}}(T_{d})$, $\mathrm{^{3/2}Fe_{i}^{+}}(C_{2v})$
and $\mathrm{^{2}Fe_{i}^{+\!+}}(T_{d})$, where the total spin/charge
is left/right-superscripted to the Fe symbol and the symmetry of the
defect is specified within parentheses. In the neutral charge state
the nearest neighboring Fe-Si distance is 2.404~Å (only 0.036~Å
longer than the bulk Si-Si bond length). The same Fe-Si distance increases
by less than 0.01~Å for each electron that is ionized from the center.
The orthorhombic distortion obtained for $\mathrm{^{3/2}Fe_{i}^{+}}$
is rather small and relates to a displacement of the Fe atom along
the $\langle100\rangle$ direction by 0.06~Å, which corresponds to
a decrease in the energy of only 50~meV with respect to the $T_{d}$
structure. The spin-1 and spin-3/2 states obtained for $\mathrm{Fe_{i}^{0}}$
and $\mathrm{Fe_{i}^{+}}$ correspond to those observed by EPR,\cite{woo60,lud60b,sie83}
and agree with previous calculations.\cite{wei96,san07} Regarding
$\mathrm{Fe_{i}^{+\!+}}$, we will discuss its potential occurrence
in a spin-2 state in p$^{+}$-type material. In this charge state
the spin-1 and spin-0 configurations are unstable --- from spin-constrained
calculations we found them to be 90~meV and 0.54~eV higher in energy
than $\mathrm{^{2}Fe_{i}^{+\!+}}$.

\begin{figure}
\includegraphics[width=8cm]{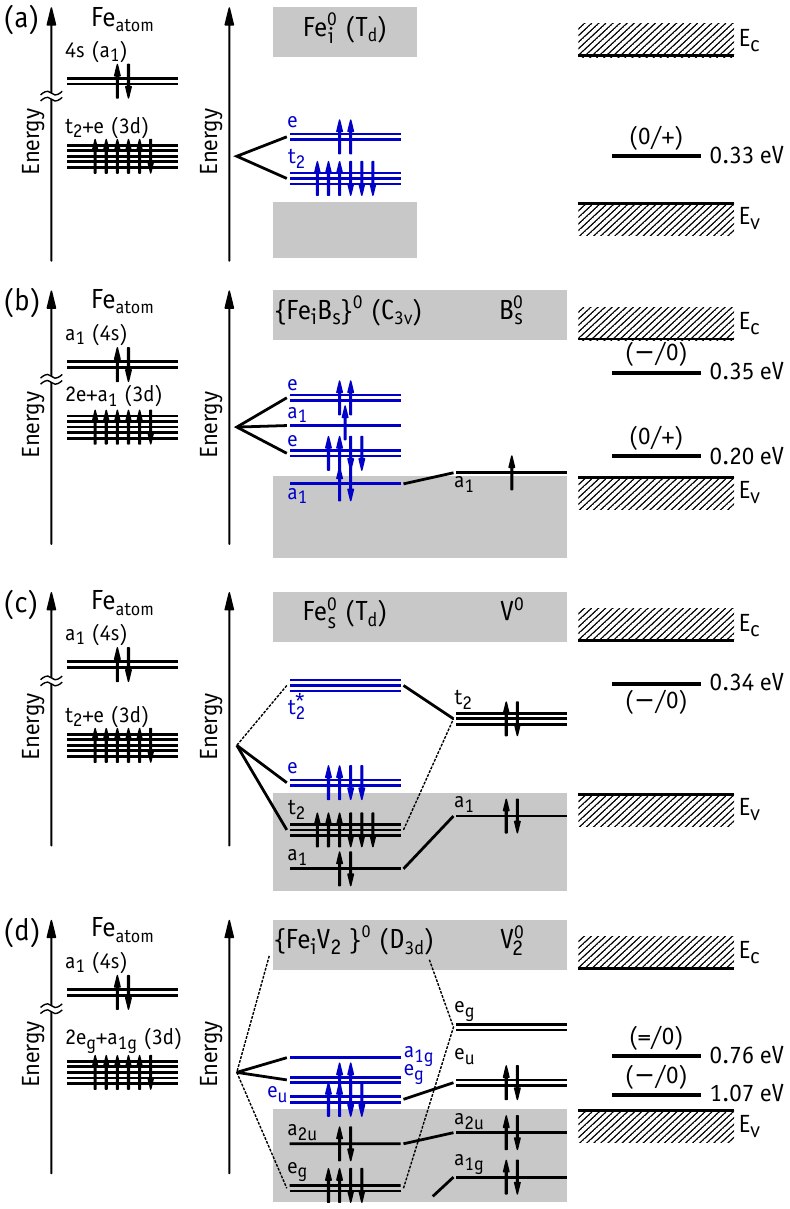}

\caption{\label{fig:levels}Electronic structure of (a) Interstitial iron,
Fe$_{\mathrm{i}}$, (b) iron-boron pair, Fe$_{\mathrm{i}}$B$_{\mathrm{s}}$,
(c) substitutional iron, Fe$_{\mathrm{s}}$, and (d) iron-divacancy,
Fe$_{\mathrm{i}}$V$_{2}$ defects in Si. Diagrams in the middle depict
the electronic coupling between atomic iron (left diagrams) and specific
defects in Si (boron, vacancy and divacancy). Diagrams on the right
show the calculated electronic levels for each case, where donor and
acceptor energies are given with respect to the valence band top and
conduction band bottom, respectively. In all diagrams, electronic
levels are labelled according to the the point group of the defect
under scrutiny. Occupied spin-up/-down states are shown as upward/downward
arrows, respectively.}
\end{figure}

From inspection of the Kohn-Sham band structure we confirm the established
model for the electronics of $\mathrm{Fe_{i}}$ in Si. The result
is depicted in Figure~\ref{fig:levels}(a) and closely follows the
model early proposed by Ludwig and Woodbury.\cite{lud60} Before we
proceed with the discussion, we note that the departure from $T_{d}$
symmetry in Fe$_{\mathrm{i}}^{+}$ is in the present context considered
small enough to justify a tetrahedral representation of its electronic
states. Basically all 3d levels of atomic Fe (in the figure labeled
within the $T_{d}$ point group symmetry as $t_{2}+e$) are found
inside the band gap of Si which, due to the tetrahedral crystal field,
are split into $t_{2}$ (low energy) and $e$ (high energy) states,
respectively. Importantly, the neutral Fe impurity adopts a non-oxidized
3d$^{8}$4s$^{0}$ configuration (unlike atomic iron which is 3d$^{6}$4s$^{2}$).
Also in line with the EPR data we found that the ionization of the
defect involves a change in the occupation of the triplet (not in
the higher energy doublet). Due to exchange interactions, electrons
occupying the $e$ state are tightly bound. This results in the formation
of a $t_{2}^{4\uparrow}+e^{\uparrow\uparrow}$ manifold with net spin
$S=3/2$. In fact, the most stable $^{1/2}$Fe$^{+}$ state was not
$t_{2}^{6}+e^{\uparrow}$, but $t_{2}^{4\downarrow}+e^{\uparrow\uparrow}$
which was 0.31~eV above the spin-3/2 ground state. The weak $C_{2v}$
distortion obtained for Fe$_{\mathrm{i}}^{+}$ suggests a possible
Jahn-Teller instability driven by the incomplete filling of the triplet
state. Experimental evidence for this effect may be inferred from
a dynamic broadening of the EPR signal of Fe$_{\mathrm{i}}^{+}$ relative
to that of Fe$_{\mathrm{i}}^{0}$.\cite{wei96} However, we note that
the calculation of Jahn-Teller symmetry breaking using methods based
on the Born-Oppenheimer approximation (as presented here), should
be considered with care -- any quantitative treatment of this effect
must account for the electron-phonon coupling. By comparing first
ionization energies of Fe and Ti impurities in 216-atom supercells
we arrive at a $\mathrm{Fe_{i}}(0/+)$ transition at $E_{\mathrm{v}}+0.33$~eV
{[}obtained from $I_{\mathrm{Fe}}(0/+)-I_{\mathrm{Ti}}(0/+)=-0.54$~eV{]}.
This compares fairly well with the experimental figure of $E_{\mathrm{v}}+0.38$~eV
(see for instance Refs.~\onlinecite{fei78} and \onlinecite{wun82}).
Similarly, by comparison with $E_{\mathrm{c}}-\mathrm{Cu_{i}}(0/+)=0.15$~eV,
the level of $\mathrm{Fe_{i}}(0/+)$ is very close to the Ti-marked
result and it is calculated at $E_{\mathrm{v}}+0.32$~eV. Since the
$\mathrm{Cu_{i}}(0/+)$ marker is referred with respect to the conduction
band minimum, a band gap width of 1.17~eV was considered in order
to calculate $\mathrm{Fe_{i}}(0/+)$ with respect to $E_{\mathrm{v}}$.

Further ionization of the defect also involved electron depletion
of the $t_{2}$ triplet state. By comparing $I_{\mathrm{Fe}}(+/+\!+)$
and $I_{\mathrm{Ti}}(+/+\!+)$ energies we were lead to a second donor
level for Fe at $E_{\mathrm{v}}+0.08$~eV. It is noted that there
is no experimental evidence for such a level. This result may suffer
from an overestimation of the stability of the high-spin $^{2}\mathrm{Fe}_{\mathrm{i}}^{+\!+}$
ground state by the use of a semi-local treatment of the exchange-correlation
potential. As a matter of fact, if we consider the energy of the $^{1}\mathrm{Fe}_{\mathrm{i}}^{+\!+}$
state, the Fe$_{\mathrm{i}}(+/+\!+)$ level becomes resonant with
the valence band.

\begin{table}
\caption{\label{tab:mossbauer-fe-si}Calculated isomer shifts, $\delta$ (mm/s)
and quadrupole splittings $\Delta$ (mm/s) for the defects under investigation
in several charge states of interest.}

\begin{tabular}{cccc}
\hline 
 & $\delta$ &  & $\Delta$\tabularnewline
\hline 
$\mathrm{Fe_{i}(0,+)}$ & 0.72, 0.67 &  & \tabularnewline
$\mathrm{FeB_{s}(-,0,+)}$ & 0.71, 0.68, 0.64 &  & $-0.04$, 0.57, 0.96\tabularnewline
$\mathrm{Fe_{s}(-,0,+)}$ & $-0.11$, $-0.13$, $-0.14$ &  & \tabularnewline
$\mathrm{FeV_{2}(-,0)}$ & 0.35, 0.39 &  & 0.40, 1.10\tabularnewline
\hline 
\end{tabular}
\end{table}

In Table~\ref{tab:mossbauer-fe-si} we find the calculated isomer
shifts and quadrupole splittings for the defects under investigation
in their relevant charge states. From the APW+lo calculations for
$\mathrm{^{1}Fe_{i}^{0}}$ we obtained a relative contact density
$\bar{n}\mathrm{(^{1}Fe_{i}^{0}})-\bar{n}(\alpha\text{-}\mathrm{Fe})=-2.725$~Bohr$^{-3}$,
and using Eq.~\ref{eq:isomershift} we arrive at an isomer shift
$\delta\mathrm{(^{1}Fe_{i}^{0}})=0.72$~mm/s. Ionization of the Fe$_{\mathrm{i}}$
impurity results in a small increase in the (absolute) contact density
that corresponds to an isomer shift of 0.67~mm/s for $\mathrm{^{3/2}Fe_{i}^{+}}$.
These results agree reasonably well with the experimental data obtained
by the ISOLDE consortium, \emph{i.e.} 0.77~mm/s and 0.72~mm/s observed
in n-type and p-type Si and assigned to $\mathrm{Fe_{i}^{0}}$ and
$\mathrm{Fe_{i}^{+}}$, respectively.\cite{gun02a} Extensive measurements
in n- and p-type samples, where $^{57}$Fe was introduced by several
methods, were also carried out by Yoshida and his group.\cite{yos06,yos07a,yos16}
Although their data for p-type Si seems to be in line with other reported
data ($\delta=0.8$~mm/s), for n-type material a resonance at $\delta=0.40$~mm/s
has been assigned to $\mathrm{Fe_{i}^{0}}$. Since our calculations
conflict with this assignment, we investigated several possible sources
of error, including (i) increasing the BZ sampling grid to MP-$4^{3}$,
(ii) bringing the 3s state of Fe into the valence to be treated within
the Kohn-Sham scheme, (iii) including an on-site correction to the
exchange-correlation energy by means of the GGA$+U+J$ approach as
proposed by Liechtenstein et~al.\cite{lie95} (with Hubbard and Hund
parameters in the range $U=2\text{-}5$~eV and $J=0\text{-}1$~eV),
and (iv) using the local density approximation to the electronic exchange-correlation
interactions. All these tests resulted in a small and negative charge-induced
shift $\delta(+/0)=-0.04(2)$~mm/s, where $\delta(+/0)\equiv\delta(\mathrm{^{3/2}Fe}_{\mathrm{i}}^{+})-\delta(\mathrm{^{1}Fe}_{\mathrm{i}}^{0})$,
which is effectively identical to that obtained from $\delta$ values
reported in Table~\ref{tab:mossbauer-fe-si}.

\begin{figure}
\includegraphics{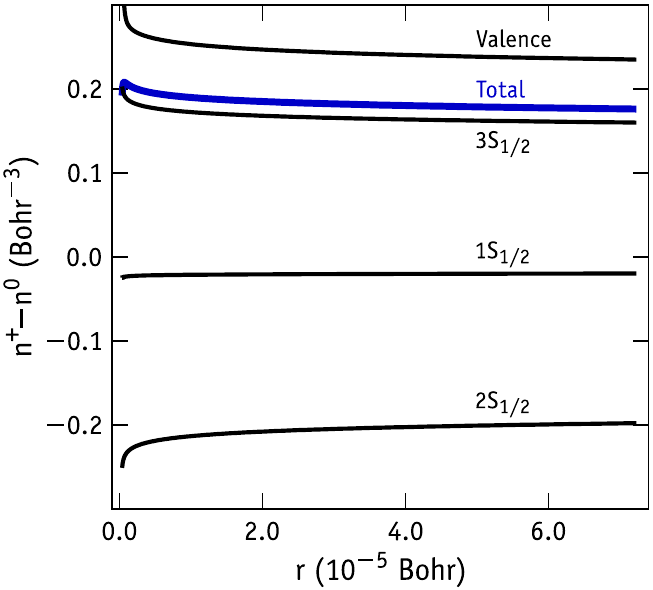}

\caption{\label{fig:rho-core-silicon}Change in the electron density upon ionization
of Fe$_{\mathrm{i}}$ in Si, $n^{+}-n^{0}$, as a function of the
radial distance from the center of the Fe nucleus. The graph depicts
the total change in the density (thick blue curve), along with partial
contributions to $n^{+}-n^{0}$ from $NS_{1/2}$ core states (up to
$N=3$) and valence states (thinner black curves). }
\end{figure}

The small calculated change in the isomer shift upon ionization of
Fe$_{\mathrm{i}}$ in Si is also in agreement with previous theoretical
reports. Early work by Katayama-Yoshida and Zunger\cite{yos85} by
means of a self-consistent spin-polarized Green's-function method
reported $\delta(+/0)=-0.06$~mm/s, and more recently Kübler and
co-workers\cite{kub93} obtained $\delta(+/0)=-0.08$~mm/s. Figure~\ref{fig:rho-core-silicon}
depicts the difference between the electron density around the $^{57}$Fe
nucleus in Fe$_{\mathrm{i}}^{+}$ and Fe$_{\mathrm{i}}^{0}$ defects,
represented as $n^{+}(r)-n^{0}(r)$. It also includes partial contributions
from S-states (where the relativistic notation is used) and valence.
Two main effects dictate the almost identical contact densities in
Fe$_{\mathrm{i}}^{+}$ and Fe$_{\mathrm{i}}^{0}$, namely (i) the
fact that the 4s state is not involved in the ionization process --
only 3d states (which are nodal at the nucleus) change occupancy,
and (ii) a considerable screening of the Fe impurity by the host crystal,
meaning that removal of one electron from the triplet state results
in a charge flow (electronic relaxation) from ligand atoms towards
the Fe site. This is demonstrated by the large positive contribution
from the valence to $n^{+}-n^{0}$. Contributions from $2S_{1/2}$
and $3S_{1/2}$ states have similar magnitudes and mutually cancel.

\subsection{Iron-boron pair in Si}

Our calculations indicate that the pairing of Fe with substitutional
B results in trigonal ($C_{3v}$) complexes with $^{2}\left\{ \mathrm{Fe_{i}B_{s}}\right\} ^{+}$,
$^{3/2}\left\{ \mathrm{Fe_{i}B_{s}}\right\} ^{0}$ and $^{1}\left\{ \mathrm{Fe_{i}B_{s}}\right\} ^{-}$
ground states. Low-spin $^{1}\left\{ \mathrm{Fe_{i}B_{s}}\right\} ^{+}$
and $^{0}\left\{ \mathrm{Fe_{i}B_{s}}\right\} ^{+}$ configurations
were 0.06~eV and 0.47~eV above the ground state. From the point
of view of its formation, it is instructive to look at the neutral
state as a Coulomb stabilized complex made of $^{3/2}\mathrm{Fe_{i}}^{+}$
next to B$_{\mathrm{s}}^{-}$. Alternatively, we may think of it as
a result of electron transfer from the high-energy d-state of Fe$_{\mathrm{i}}$
to the low-energy acceptor state of boron, meaning that the electronic
structure of $\left\{ \mathrm{Fe_{i}B_{s}}\right\} ^{q}$ resembles
that of Fe$_{\mathrm{i}}^{q+1}$ (perturbed by a B$_{\mathrm{s}}^{-}$
anion).\cite{zha01,mac05} According to Zhao \emph{et~al.}\cite{zha01}
the effect of a negatively charged B$_{\mathrm{s}}^{-}$ next to Fe$_{\mathrm{i}}$
is to raise the energy of the Fe$_{\mathrm{i}}$ levels due to Coulomb
repulsion. The magnitude of this interaction has been estimated from
the measured Fe$_{\mathrm{i}}$B$_{\mathrm{s}}$ and Fe$_{\mathrm{i}}$
levels\cite{fei78,lem81} as $\mathrm{Fe_{i}B_{s}}(-/0)-\mathrm{Fe_{i}}(0/+)=(E_{\mathrm{c}}-0.26)-(E_{\mathrm{v}}+0.38)=$
0.53~eV, where the band gap of Si at $T=0$~K is considered to be
$E_{\mathrm{c}}-E_{\mathrm{v}}=1.17$~eV. From here, and hypothesizing
that the $\mathrm{Fe_{i}B_{s}}(0/+)$ level also originates from a
Coulomb-raised $\mathrm{Fe_{i}}(+/+\!+)$ transition resonant with
the Si valence band, we readily arrive at an estimate for its location
as $\mathrm{Fe_{i}}(+/+\!+)=\mathrm{FeB_{s}}(0/+)-0.53=E_{\mathrm{v}}-0.42$~eV,
where again we made use of the measured donor level of Fe$_{\mathrm{i}}$B$_{\mathrm{s}}$
at $E_{\mathrm{v}}+0.11$~eV.\cite{bro85} Obviously this analysis
is based on the premise that the strength of the interaction between
B$_{\mathrm{s}}^{-}$ and the two levels of Fe$_{\mathrm{i}}$ is
the same, which although reasonable, is yet to be demonstrated.

The calculated electronic structure of neutral Fe$_{\mathrm{i}}$B$_{\mathrm{s}}$
is depicted in Figure~\ref{fig:levels}(b). Analogously to Fe$_{\mathrm{i}}$,
we found that the ionization/electron trapping of/by Fe$_{\mathrm{i}}$B$_{\mathrm{s}}$
essentially involves a change in the occupation of a $a_{1}+e$ manifold
(once a $t_{2}$ triplet localized on Fe$_{\mathrm{i}}$) below a
$e^{\uparrow\uparrow}$ spin-1 doublet. The calculated electronic
levels of Fe$_{\mathrm{i}}$B$_{\mathrm{s}}$ marked with Ti are located
at Fe$_{\mathrm{i}}$B$_{\mathrm{s}}(0/+)=E_{\mathrm{v}}+0.20$~eV
and Fe$_{\mathrm{i}}$B$_{\mathrm{s}}(-/0)=E_{\mathrm{c}}-0.35$~eV,
in fair agreement with the corresponding experimental levels measured
at $E_{\mathrm{v}}+0.11$~eV and $E_{\mathrm{c}}-0.26$~eV, respectively.\cite{nil93}
Analogous calculations using $\mathrm{Cu_{i}}(0/+)$ and $\mathrm{Cu_{s}}(-/0)$
markers gave Fe$_{\mathrm{i}}$B$_{\mathrm{s}}(0/+)=E_{\mathrm{v}}+0.19$~eV
and Fe$_{\mathrm{i}}$B$_{\mathrm{s}}(-/0)=E_{\mathrm{c}}-0.48$~eV.
The latter figure is edging the usual error bar of the marker method,
perhaps because a substitutional marker is not the best choice to
compare with an interstitial defect. The fact that we place the Fe$_{\mathrm{i}}$B$_{\mathrm{s}}(0/+)$
level about 0.1~eV above the measurements suggests that the calculated
Fe$_{\mathrm{i}}(+/\!+\!+)$ transition reported above at $E_{\mathrm{c}}+0.08$~eV
is spurious.

Likewise, for the electronic structure, the calculated isomer shifts
for $\left\{ \mathrm{Fe_{i}B_{s}}\right\} ^{q}$ are very similar
to those for Fe$_{\mathrm{i}}^{q+1}$. The results are shown in Table~\ref{tab:mossbauer-fe-si},
where calculated $\delta$ values of 0.71~mm/s and 0.68~mm/s for
$\left\{ \mathrm{FeB_{s}}\right\} ^{-}$ and $\left\{ \mathrm{FeB_{s}}\right\} ^{0}$
are very close to 0.72~mm/s and 0.67~mm/s measured for Fe$_{\mathrm{i}}^{0}$
and Fe$_{\mathrm{i}}^{+}$, respectively. The fact that we are now
dealing with a trigonal center implies the existence of a non-zero
EFG at the Fe nucleus and a quadrupole splitting in the Mössbauer
signal. The calculations (also reported in Table~\ref{tab:mossbauer-fe-si})
indicate splittings of $\Delta=-0.04$~mm/s, 0.57~mm/s and 0.96~mm/s
for $\left\{ \mathrm{FeB_{s}}\right\} ^{-}$ and $\left\{ \mathrm{FeB_{s}}\right\} ^{0}$
and $\left\{ \mathrm{FeB_{s}}\right\} ^{+}$. These results indicate
that the nuclei of Fe ions in neutral $\left\{ \mathrm{Fe^{+}B_{s}^{-}}\right\} $
and positively charged $\left\{ \mathrm{Fe^{+\!+}B_{s}^{-}}\right\} $
complexes experience much larger EFGs than neutral Fe in negatively
charged $\left\{ \mathrm{Fe^{0}B_{s}^{-}}\right\} $, where there
is no Coulomb attraction between the impurity pair. These results
differ considerably from Mössbauer data acquired at $T\approx900$~K
where a signal with a centroid velocity $\delta=0.93$~mm/s and line
splitting $\Delta=1.6$~mm/s was connected to the Fe$_{\mathrm{i}}$B$_{\mathrm{s}}$
pair (presumably in the neutral charge state).\cite{gun06b} They
are also at variance with a recent proposal that the Fe$_{\mathrm{i}}$B$_{\mathrm{s}}$
pair gives rise to a single resonant line (presumably without or with
unresolvable quadrupole splitting) at $\delta=0.42$~mm/s.\cite{yos16}
On the other hand, they are in much better agreement with earlier
work by Bergholz on $^{57}$Co-contaminated p-type Si, where at least
three doublets were observed with intensities that depended on the
temperature of the measurements, and were all assigned to $^{57}$Co$_{\mathrm{i}}$B$_{\mathrm{s}}$
pairs. All these doublets had $\delta$ values of about 0.7~mm/s,
and were split by either 0.21, 0.54 and 1.04~mm/s, the one with larger
$\Delta$ being assigned to the $\langle111\rangle$-oriented structure
studied here.\cite{ber83} Such mixed agreement calls for further
theoretical analysis and experiments.

\subsection{Substitutional iron in silicon}

Substitutional Fe was studied in the neutral and negative charge states.
We obtained ground state configurations $^{0}\mathrm{Fe_{s}^{0}}(T_{d})$
and $^{1/2}\mathrm{Fe_{s}^{-}}(C_{2v})$. The tetrahedral structure
had 2.255~Å Fe-Si bonds, while the Fe atom in the $C_{2v}$ structure
was displaced from the perfect lattice by 0.11~Å along $\langle001\rangle$,
resulting in Fe-Si bonds either 2.249~Å or 2.239~Å long. These are
somewhat shorter than Si-Si bonds in bulk Si (2.368~Å) and also shorter
than Fe-Si bonds in $\beta$-FeSi$_{2}$ (2.33-2.42~Å). The perfect
tetrahedral $^{1/2}\mathrm{Fe_{s}^{-}}(T_{d})$ defect is unstable
-- a symmetry-constrained relaxation of the structure gave a total
energy 0.11~eV above $^{1/2}\mathrm{Fe_{s}^{-}}(C_{2v})$. Considering
the formation enthalpy difference of Fe$_{\mathrm{i}}$ to Fe$_{\mathrm{s}}$
($-0.5$~eV), combined with a calculated formation enthalpy of 3.60~eV
for a neutral Si vacancy (V), we estimate a binding energy between
Fe$_{\mathrm{s}}$ and V of about 3~eV. This means that if the concentration
of vacancies is above the equilibrium level, (\emph{ex.} after electron
irradiation), and if the temperature is such that the vacancies are
mobile, a considerable fraction of Fe$_{\mathrm{i}}$ will readily
become substitutional, unless there is a high enough barrier or a
more efficient trapping center for vacancies preventing the $\mathrm{Fe_{i}}+\mathrm{V}\rightarrow\mathrm{Fe_{s}}$
reaction from occurring. We will come back to this issue below.

The electronic structure of $\mathrm{Fe_{s}}$ in Si is represented
in Figure~\ref{fig:levels}(c), and it is well described by the Watkins
vacancy model.\cite{wat92} Accordingly, it results from the resonance
between $t_{2}$ levels from the iron $3d$ manifold and the $t_{2}$
levels from the Si vacancy. The result is the formation of a fully
occupied $t_{2}^{6}$ bonding state (in the valence) along with an
empty $t_{2}^{*}$ anti-bonding counterpart deep in the gap. The fully
occupied $e$-component from the $3d$ manifold of Fe is edging the
valence band top. We note that (i) like the $t_{2}$ gap state in
Fe$_{\mathrm{i}}$, the $t_{2}^{*}$ state of Fe$_{\mathrm{s}}$ is
nodal on the Fe atom, but (ii) unlike in Fe$_{\mathrm{i}}$, $t_{2}^{*}$
is reminiscent of the vacancy states and it is strongly localized
on Si atoms.

Early calculations based on the Green's function method anticipated
an inert Fe$_{\mathrm{s}}$ center with no levels in the gap,\cite{zun82,bee85}
\emph{i.e.}, the $e$ and $t_{2}^{*}$ states were predicted to lie
below $E_{\mathrm{v}}$ and above $E_{\mathrm{c}}$, respectively.
More recent density functional calculations placed the $t_{2}^{*}$
state well within the gap, and calculated an acceptor level either
at $E_{\mathrm{c}}-0.41$~eV or $E_{\mathrm{c}}-0.29$~eV, depending
on the calculation specifics, such as the type of pseudopotentials
and basis functions employed.\cite{est08} These results challenge
both theorists and experimentalists, suggesting that further studies
should be carried out in order to identify the Fe$_{\mathrm{s}}$
defect in Si. By comparing the electron affinity of Fe$_{\mathrm{s}}$
and Ti$_{\mathrm{i}}$ defects we arrive at an acceptor level for
Fe$_{\mathrm{s}}$ at $E_{\mathrm{c}}-0.20$~eV, in line with Estreicher
and co-workers\cite{est08} who using a methodology much similar to
ours obtained Fe$_{\mathrm{s}}(-/0)$ at $E_{\mathrm{c}}-0.29$~eV.
Experimentally, there is not much data in the literature regarding
the electrical levels of substitutional Fe. Perhaps the work by Kaminski
et~al.\cite{kam03} provide us what it could be a signature for this
defect. Accordingly, a trap at $E_{\mathrm{c}}-0.38$~eV was assigned
to an acceptor transition of Fe$_{\mathrm{s}}$, although this link
was solely based on annealing and concentration arguments.\cite{tan95,kam03}
Our calculated Fe$_{\mathrm{s}}(-/0)$ transition is about 0.2~eV
off the aforementioned trap, which is about the expected error of
the marker method if the electronic structure of the marker is far
from that of the defect under scrutiny. Here we are comparing an acceptor
transition of a defect which very much resembles the vacancy electronic
structure (Fe$_{\mathrm{s}}$) with an another acceptor transition
involving electron capture at the 3d level of interstitial Ti. However,
comparing electron affinities of $\mathrm{Cu_{s}(-/0)}$ and $\mathrm{Cu_{s}(=/-)}$
with those of $\mathrm{Fe{}_{s}(-/0)}$ and $\mathrm{Fe{}_{s}(=/-)}$
we obtain levels at $E_{\mathrm{c}}-0.34$~eV (close to Kaminski's
trap at $E_{\mathrm{c}}-0.38$~eV)\cite{kam03} and $E_{\mathrm{c}}+0.11$~eV
respectively, indicating that $\mathrm{Fe{}_{s}}$ only has one acceptor
level within the gap. While these results are consistent with those
obtained using the $\mathrm{Ti_{i}}(-/0)$ marker, due to similarities
between $\mathrm{Cu{}_{s}}$ and $\mathrm{Fe{}_{s}}$ centers we anticipate
that the $\mathrm{Cu{}_{s}}$\nobreakdash-marked $\mathrm{Fe{}_{s}}(-/0)$
level is likely to be our best estimate.

The calculated isomer shifts of neutral and negatively charged Fe$_{\mathrm{s}}$
are $-0.13$~mm/s and $-0.11$~mm/s, respectively. Their magnitude
is slightly overestimated with respect to the measurements, but unlike
previous calculations,\cite{kub93} we obtain the right sign, meaning
that the contact density on Fe$_{\mathrm{s}}$ is slightly higher
than in $\alpha$-Fe, most probably due to the additional electrons
from the $a_{1}$ vacancy state (see Fig.~\ref{fig:levels}(c)).
Notably, like in Fe$_{\mathrm{i}}$, the charge state dependence of
the isomer shift is rather small -- the capture of an electron by
Fe$_{\mathrm{s}}^{0}$ changes $\delta$ by only 0.02~mm/s, again
due to the fact that the 4s state of iron in Fe$_{\mathrm{s}}$ is
empty (the oxidation state is 0) and the $t_{2}^{*}$ acceptor state
is nodal on Fe.

\subsubsection{Iron-vacancy pair}

We investigated the Fe$_{\mathrm{i}}$V pair by first looking at its
ground state electronic structure, and then by inspecting its stability
against transformation to substitutional Fe, \emph{i.e.} by following
the reaction $\mathrm{Fe_{i}}+\mathrm{V}\rightarrow\mathrm{Fe_{s}}$.
The lowest energy structure of Fe$_{\mathrm{i}}$V, made of separate
Fe$_{\mathrm{i}}$ and V defects, was that of a Fe$_{\mathrm{i}}$
with a missing Si nearest neighbor along $\langle111\rangle$. Ground
states were all low-spin $^{1/2}\{\mathrm{Fe_{i}V}\}^{+}$, $^{0}\{\mathrm{Fe_{i}V}\}^{0}$
and $^{1/2}\{\mathrm{Fe_{i}V}\}^{-}$, while higher-spin states $^{3/2}\{\mathrm{Fe_{i}V}\}^{+}$,
$^{1}\{\mathrm{Fe_{i}V}\}^{0}$ and $^{3/2}\{\mathrm{Fe_{i}V}\}^{-}$
where metastable by 0.17~eV, 0.21~eV and 0.45~eV, respectively.
Placing the vacancy at the second nearest neighboring site along $\langle100\rangle$
to the Fe atom resulted in $\{\mathrm{Fe_{i}}\text{-Si-V}\}^{+,0,-}$
structures with spin 1/2, 0 and 1/2 and relative energy 1.18~eV,
1.33~eV and 1.47~eV above their respective Fe$_{\mathrm{i}}$V ground
states (and respective charge states). These figures are still low
when compared with the $\sim3$~eV binding energy of Fe$_{\mathrm{s}}$
from infinitely separated V and Fe$_{\mathrm{i}}$ defects (see above).

In order to follow the energetics of the $\mathrm{Fe_{i}}+\mathrm{V}\rightarrow\mathrm{Fe_{s}}$
reaction we employed the nudged elastic band method\cite{hen00a,hen00b}
on a two-step process, namely (1) $\mathrm{Fe_{i}}\text{-}\mathrm{Si}\text{-}\mathrm{V}\rightarrow\mathrm{Fe_{i}}\mathrm{V}$
and (2) $\mathrm{Fe_{i}}\mathrm{V}\rightarrow\mathrm{Fe_{s}}$. A
total of 9 intermediate structures (images) were considered between
the reactant/product structures from each step. We assume that in
the presence of vacancies, the eventual formation of $\{\mathrm{Fe_{i}}\text{-Si-V}\}$
is limited by an activation energy that corresponds to the migration
barrier of the vacancy. This barrier depends on the charge state of
the traveling V defect and was accurately measured by Watkins as 0.33~eV,
0.45~eV and 0.18~eV for V$^{+\!+}$, V$^{0}$ and V$^{=}$,\cite{wat92}
respectively. Such low values imply that $\{\mathrm{Fe_{i}}\text{-Si-V}\}$
should form before Fe$_{\mathrm{i}}$ becomes mobile, even below room
temperature. For reaction (1) we obtained barriers of 0.2~eV, 0.18~eV
and 0.18~eV for positively charged, neutral and negatively charged
supercells (spin 1/2, 0 and 1/2), respectively. For step (2) we obtain
barriers of 0.17~eV, 0.18~eV and 0.15~eV, again for charge states
$+$, $0$ and $-$, respectively (spin 1/2, 0 and 1/2). These results
are in apparent disagreement with those from Ref.~\onlinecite{est08}
where a barrier above 0.45~eV was reported. However we note we are
comparing two different mechanisms. In Ref.~\onlinecite{est08} a
series of molecular dynamics runs were performed in which the Si atoms
in the supercell were allowed to relax but the Fe atom was forced
to move at constant speed along a trigonal axis, starting at the second
neighboring tetrahedral interstitial site (from V) over the hexagonal
site to the nearest neighboring T site, and finally into the vacancy
site. In our opinion, this approach unjustifiably assumes that the
moving defect is Fe (while V is assumed to be static). 

Our predicted barriers are not compatible with a Fe$_{\mathrm{i}}$V
complex stable up to 500$^{\circ}\text{C}$ as reported in Ref.~\onlinecite{mch02}
and references therein, neither support the early assignment of the
EPR NL19 signal to Fe$_{\mathrm{i}}$V, which was then shown to be
stable at least up to 160$^{\circ}\text{C}$.\cite{mul82} Although
the $^{3/2}\{\mathrm{Fe_{i}}\text{V}\}^{+}$ candidate is only 0.17~eV
less stable than $^{1/2}\{\text{Fe-V}\}^{+}$, we did not find a stable
alternative spin-3/2 Fe$_{\mathrm{i}}$V defect that could be connected
to NL19. Based on the above results, our view is that Fe$_{\mathrm{i}}$V
will not survive at room temperature and above, and that NL19 could
be related to a more stable vacancy-Fe$_{\mathrm{i}}$ complex (also
involving a single Fe atom), with $^{3/2}$Fe$_{\mathrm{i}}$V$_{2}^{-}$
being a likely candidate (see below).

\subsection{Iron-divacancy pair in Si}

Here we focus mostly on the stable iron-divacancy (Fe$_{\mathrm{i}}$V$_{2}$)
structure which, as reported in Ref.~\onlinecite{est08}, consists
on a Fe atom right at the center of a divacancy (at the bond-center
site of the otherwise perfect Si lattice). Accordingly, we obtain
$^{1/2}$Fe$_{\mathrm{i}}$V$_{2}^{+}$, $^{1}$Fe$_{\mathrm{i}}$V$_{2}^{0}$,
$^{3/2}$Fe$_{\mathrm{i}}$V$_{2}^{-}$ and $^{1}$Fe$_{\mathrm{i}}$V$_{2}^{=}$,
all trigonal structures with $D_{3d}$ symmetry. Alternative spin
configurations $^{3/2}$Fe$_{\mathrm{i}}$V$_{2}^{+}$, $^{2}$Fe$_{\mathrm{i}}$V$_{2}^{0}$,
$^{1/2}$Fe$_{\mathrm{i}}$V$_{2}^{-}$ and $^{0}$Fe$_{\mathrm{i}}$V$_{2}^{=}$
where found less stable by 0.02~eV, 0.03~eV, 0.12~eV and 0.37~eV,
respectively.

As depicted in Figure~\ref{fig:levels}(d), the electronic structure
of Fe$_{\mathrm{i}}$V$_{2}$ results from the overlap of the 3d states
of the Fe atom ($2e_{g}+a_{1g}$ in a $D_{3d}$ representation) with
the $e_{g}$ and $e_{u}$ state of the Si divacancy. The $e_{g}$
states from both defects mix strongly, resulting in bonding and anti-bonding
levels in the valence and conduction bands, respectively. The formation
energy of a neutral Fe$_{\mathrm{i}}$V$_{2}$ is 5.47~eV, which
along with the calculated formation energies for Fe$_{\mathrm{s}}$
and V gives a binding energy of 1.33~eV for the reaction $\mathrm{Fe_{s}}+\mathrm{V}\rightarrow\mathrm{Fe_{i}V_{2}}$.
This figure is not far from the 1.56~eV previously found by the Estreicher
group,\cite{est08} and suggests that this is a \emph{strong} defect
that can survive well above room temperature.

Other structures for Fe$_{\mathrm{i}}$V$_{2}$ were investigated.
The asymmetric structure with the Fe atom located at one of the vacant
sites ($\mathrm{Fe_{s}V}$ structure) was unstable -- upon relaxation,
the Fe atom moved without an impeding barrier to the bond center site
between the two vacancies. Other structures like $\mathrm{Fe_{s}\text{-}Si\text{-}V}$
or $\mathrm{Fe_{s}\text{-}Si\text{-}Si\text{-}V}$, where a vacancy
is located at the second and third neighboring site to Fe$_{\mathrm{s}}$
were found to be 0.98~eV and 1.26~eV above the ground state.

Also in agreement with the authors of Ref.~\onlinecite{est08} we
found that Fe$_{\mathrm{i}}$V$_{2}$ (with $D_{3d}$ symmetry) is
not a donor but rather a multiple acceptor. While they report a $(-/0)$
level at $E_{\mathrm{c}}-0.73$~eV or $E_{\mathrm{c}}-0.64$~eV
(depending on the method specifics), we obtain a first acceptor level
at $E_{\mathrm{c}}-0.92$~eV using the Ti$_{\mathrm{i}}$-marker.
The use of the more suitable Cu$_{\mathrm{s}}$ marker gives $\mathrm{Fe_{i}V_{2}}(-/0)=E_{\mathrm{v}}+0.10$~eV
and $\mathrm{Fe_{i}V_{2}}(=/-)=E_{\mathrm{v}}+0.41$~eV. Recently,
Tang and co-workers\cite{tan13} reported electrical measurements
in p-type electron-irradiated Si contaminated with Fe. From a series
of 30-min isochronal anneals they found that above $T\approx150^{\circ}$C
mobile Fe$_{\mathrm{i}}$ could interact with divacancies to form
a hole trap at $E_{\mathrm{v}}+0.29$~eV (labeled H29) which was
connected to a metastable Fe$_{\mathrm{i}}$-V$_{2}$ complex. By
rising the annealing temperature above $T\approx200^{\circ}$C the
H29 signal transformed to a more stable and deeper hole trap (H34)
at $E_{\mathrm{v}}+0.34$~eV, which was assigned to the $D_{3d}$-symmetric
Fe$_{\mathrm{i}}$-V$_{2}$ defect. While its was not possible to
establish the charge states involved in the H34 transition, our calculations
strongly suggest that it relates to $\mathrm{Fe_{i}V_{2}}(=/-)$.

The calculated Mössbauer parameters of Fe$_{\mathrm{i}}$V$_{2}$
are reported in Table~\ref{tab:mossbauer-fe-si}. Being a non-cubic
center, it should give rise to a quadrupole splitting in the MS signal.
Accordingly, for the neutral and negatively charged defects we anticipate
resonance speeds of $\delta=0.35$~mm/s and $\delta=0.39$~mm/s
for the centroid, and splittings $\Delta=1.1$~mm/s and $\Delta=0.4$~mm/s,
respectively. Analogously to previous Fe-related defects, the change
of $\delta$ values upon charge state transition is small. On the
other hand, the quadrupole splitting decreases by more that 50\% upon
electron capture by $\{\mathrm{Fe_{\mathrm{i}}V}\}^{0}$. This effect
comes from the occupation of an $a_{1g}$ axial state centered on
the Fe atom. We note that $\{\mathrm{Fe_{\mathrm{i}}V}\}^{-}$, which
is expected to be particularly stable in high-resistivity p-type and
intrinsic material, should give rise to resonances close to $0.8$~mm/s
and $-0.1$~mm/s. These coincide with the isomer shifts of Fe$_{\mathrm{i}}$
and Fe$_{\mathrm{s}}$, which may mask (or even undermine) the analysis
of the experimental data, particularly in the case where ion implantation
has been utilized to introduce the $^{57}$Fe species.

\section{Iron in other group-IV semiconductors\label{sec:other-group-iv}}

Mössbauer parameters were also calculated for elemental $\mathrm{Fe_{s}}$
and $\mathrm{Fe_{\mathrm{i}}}$ defects in other group-IV materials,
namely in Ge, Diamond, 3C-SiC (cubic phase) and in Si-rich SiGe alloys.
Like in Si, the defects were found to minimize the total energy in
the $^{0}\mathrm{Fe_{s}^{0}}(T_{d})$ and $^{1}\mathrm{Fe_{\mathrm{i}}^{0}}(T_{d})$
states. The distance between Fe and its nearest neighbors expanded
relative to the unrelaxed values, except for $\mathrm{Fe_{\mathrm{s}}}$
in Si and Ge (where it contracted) and $\mathrm{Fe_{\mathrm{C}}}$
in 3C-SiC (where it remained effectively unaltered).

\begin{figure}
\includegraphics[width=1\columnwidth]{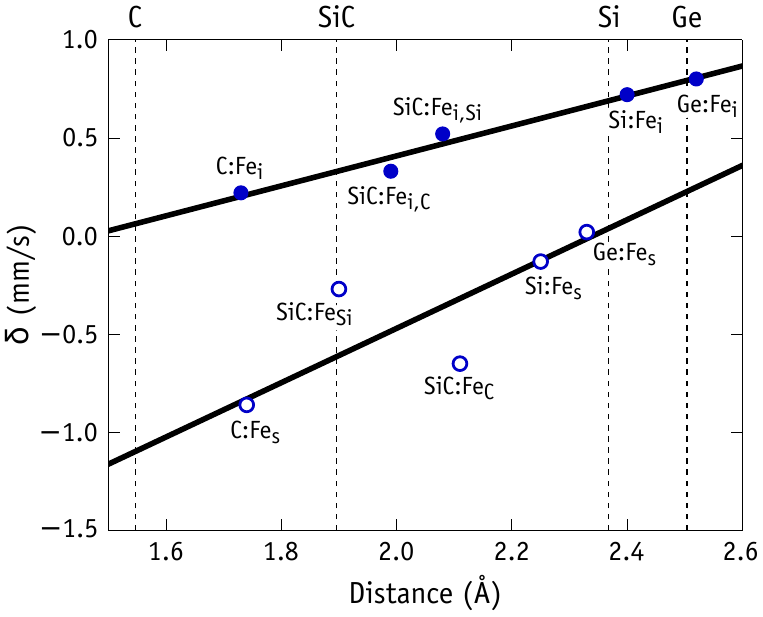}

\caption{\label{fig:is-vs-bond}IS as a function of the distance between Fe
and its first neighbors for $\mathrm{Fe_{s}}$ and $\mathrm{Fe_{\mathrm{i}}}$
defects in Diamond, Ge, Si and 3C-SiC. Vertical dashed lines indicate
the bond lengths in the corresponding bulk semiconductors.}
\end{figure}

Overall, there is good agreement between the isomer shifts observed
in Ge, Diamond and SiC, collected in Table~\ref{tab:ms-survey},
and those from our calculations reported in Table~\ref{tab:other-group-4}.
As depicted in Figure~\ref{fig:is-vs-bond} and in agreement with
Ref.~\onlinecite{gun06}, we found an approximate linear trend between
IS values and the distance between Fe and its first neighbors. IS
values for $\mathrm{Fe_{\mathrm{i}}}$ in Ge, Diamond and $\mathrm{Fe_{\mathrm{i,C}}}$
in SiC match the experimental results, and for all other IS values
the agreement is within 0.04~mm/s and 0.14~mm/s. It is noted that,
while the IS obtained for $\mathrm{Fe_{\mathrm{s}}}$ in Diamond deviates
markedly from previous calculations,\cite{bha98} the difference between
the calculation methods most likely accounts for this discrepancy.
Notably, the relaxed $\mathrm{Fe_{\mathrm{i}}}$-C and $\mathrm{Fe_{s}}$-C
distances in Diamond were effectively equal, yet $\mathrm{\delta(Fe_{\mathrm{i}})}$
and $\mathrm{\delta(Fe_{s})}$ are 1.08~mm/s apart. However, as we
move towards heavier elements, the relative differences between $\mathrm{\delta(Fe_{\mathrm{i}})}$
and $\mathrm{\delta(Fe_{s})}$ decrease with increasing distance between
Fe and its neighbors, to a minimum value of 0.76~mm/s calculated
for Ge.

To the best of our knowledge, $\delta(\mathrm{Fe_{C})}$ in SiC has
not been determined experimentally. We obtain $\delta(\mathrm{Fe_{C})=-0.65}$~mm/s,
which is about the same value previously calculated for Fe$_{\mathrm{C}}$
in 6H-SiC.\cite{elz14} We note though, that experimental IS resonances
for substitutional and interstitial Fe in both 6H-SiC and 3C-SiC follow
similar trends.\cite{bha08} Following the methodology outlined in
Section~\ref{sub:formation-energy}, formation energy calculations
for $\mathrm{Fe_{Si}}$ and $\mathrm{Fe_{C}}$ in 3C-SiC cubic supercells,
yielded $E_{\mathrm{f}}\mathrm{(Fe_{Si})}=$~2.99~eV, 2.72~eV and
3.26~eV, and $E_{\mathrm{f}}\mathrm{(Fe_{C})}=$~4.78~eV, 5.05~eV
and 4.51~eV under stoichiometric, Si-rich and C-rich crystal growth
conditions respectively. As $E_{\mathrm{f}}\mathrm{(Fe_{Si})}$ is
between 1.2~eV and 2.3~eV lower than $E_{\mathrm{f}}\mathrm{(Fe_{C})}$,
it suggests a preferential incorporation of $\mathrm{Fe}$ in the
Si sites, allowing us to assign the observed peak at $\delta=-0.23$~mm/s
reported in Ref.~\onlinecite{bha08} to $\mathrm{Fe_{Si}}$ (compared
to the calculated value of $-0.27$~mm/s).

Formation energies were also calculated for interstitial iron in SiC,
where $E_{\mathrm{f}}\mathrm{(Fe_{i,C})}=4.85$~eV and $E_{\mathrm{f}}\mathrm{(Fe_{i,Si})}=5.74$~eV,
indicating a preference for sites with first-neighboring C atoms.
This result explains the transformation of the Fe$_{\mathrm{i,2}}$
peak to the Fe$_{\mathrm{i,1}}$ peak observed in Ref.~\onlinecite{bha08}
upon annealing the SiC samples, allowing us to assign Fe$_{\mathrm{i,1}}$
and Fe$_{\mathrm{i,2}}$ peaks to $\mathrm{Fe_{i,C}}$ and $\mathrm{Fe_{i,Si}}$,
respectively.

For Si-rich SiGe alloys, the Mössbauer parameters were studied as
a function of the distance between a $\mathrm{Fe_{s}}$ ($\mathrm{Fe_{\mathrm{i}}}$)
defect and a neighboring substitutional Ge atom (from first to fourth
neighboring distance). The presence of the Ge atom produced a deformation
of the local electron density in all cases. The values of the IS and
QS for $\mathrm{Fe_{\mathrm{i}}}$ and $\mathrm{Fe_{s}}$ decrease
monotonically with the distance between the Fe impurity and the Ge
atom, and both $\mathrm{\delta(Fe_{\mathrm{i}})}$ and $\mathrm{\delta(Fe_{s})}$
are very similar to values calculated in bulk Si. Interestingly, the
QS for $\mathrm{Fe_{i}}$ with a Ge first neighbor is 1.25~mm/s,
about seven times the value obtained with a Ge fourth neighbor. However,
the relative energies obtained from PAW calculations for second, third
and fourth neighboring Fe$_{\mathrm{i}}$-Ge pairs are $-0.12$~eV,
$-0.17$~eV and $-0.17$~eV, with respect to the first neighboring
pair, allowing us to conclude that Fe$_{\mathrm{i}}$ atoms will effectively
avoid Ge atoms in the alloy. Hence, the Mössbauer spectrum of Fe$_{\mathrm{i}}$
in Si-rich SiGe alloys is expected to be analogous to the signal in
bulk Si, although broadened by quadrupole splittings of remote Fe$_{\mathrm{i}}$-Ge
pairs and internal strain fields. For Fe$_{\mathrm{s}}$, structures
with Ge atoms at first to fourth neighboring sites have essentially
the same small relative energies (within 0.03~eV), suggesting that
in principle there will be no preferential distribution of Fe$_{\mathrm{s}}$
with respect to the location of the Ge minority species. However,
due to a sizable QS for Ge at first and second nearest neighboring
positions, considerable differences are expected in the corresponding
Mössbauer spectra with respect to the bulk Si measurements. Obviously
these deviations will depend on the fraction of the Ge concentration
with respect to that of Si.

\begin{table}
\caption{\label{tab:other-group-4}Calculated IS ($\delta$ in mm/s) and QS
($\Delta$ in mm/s) values for $\mathrm{Fe_{\mathrm{i}}}$ and $\mathrm{Fe_{\mathrm{s}}}$
defects in Ge, C, 3C-SiC and SiGe alloys. For the latter material,
numbered rows 1, 2, 3 and 4 refer to first, second, third and fourth
neighboring Fe-Ge pairs.}

\begin{tabular}{cccccc}
\hline 
 &  & $\delta(\mathrm{Fe_{i}})$ & $\delta(\mathrm{Fe_{s}})$ & $\Delta(\mathrm{Fe_{i}})$ & $\Delta(\mathrm{Fe_{s}})$\tabularnewline
\hline 
C &  & 0.22 & $-0.86$ &  & \tabularnewline
 &  &  &  &  & \tabularnewline
SiC & (Fe$_{\mathrm{i,Si}}$/Fe$_{\mathrm{C}}$) & 0.52 & $-0.27$ &  & \tabularnewline
 & (Fe$_{\mathrm{i,C}}$/Fe$_{\mathrm{Si}}$) & 0.33 & $-0.65$ &  & \tabularnewline
 &  &  &  &  & \tabularnewline
SiGe & 1 & 0.71 & $-0.13$ & 1.25 & 1.55\tabularnewline
 & 2 & 0.69 & $-0.17$ & 0.19 & 0.42\tabularnewline
 & 3 & 0.69 & $-0.16$ & 0.19 & 0.10\tabularnewline
 & 4 & 0.69 & $-0.16$ & 0.17 & 0.01\tabularnewline
 &  &  &  &  & \tabularnewline
Ge &  & 0.80 & 0.02 &  & \tabularnewline
\hline 
\end{tabular}
\end{table}

\section{Conclusions\label{sec:conclusions}}

In this work we combined pseudopotential and all-electron density
functional calculations of Fe-related defects in group-IV semiconductors
(mostly in Si, but also in Ge, C, SiC and Si-rich SiGe alloys). Our
aim was to investigate electronic and electron-nuclear coupling properties,
and compare them to those measured with several spectroscopic techniques,
Mössbauer Spectroscopy in particular.

After a short review of previous experimental and theoretical reports
on $^{57}$Fe Mössbauer parameters for Fe-related defects in group-IV
semiconductors, we described the theoretical methodologies employed
by us. This includes methods to obtain defect structures, total energies
and electron densities (PAW and APW+lo methods), to calculate isomer
shifts and quadrupole splittings, formation energies and charge transition
levels.

We provided a detailed description of the calculation of the Mössbauer
calibration constants, $\alpha$ and $Q$. These quantities allowed
us to obtain the isomer shifts and quadrupole splittings from the
contact densities and electric field gradients calculated from first-principles.
To this end, we calculated relative contact densities and EFG values
for Fe in a comprehensive set of Fe-related compounds, and established
linear relations with corresponding IS and QS values obtained experimentally.
The resulting values $\alpha=0.26$~Bohr$^{3}$\,mm/s and $Q=0.17$~b
agree well with previous calculations reported in the literature.

We devoted Section~\ref{sec:iron-defects-in-si} to the study of
Fe defects in Si. We started by looking at the relative stability
and upper concentration limit of interstitial iron relative to that
of substitutional iron in Si (assuming equilibrium conditions across
a Si/$\beta$-FeSi$_{2}$ interface). We found an enthalpy of formation
for Fe$_{\mathrm{i}}$ of 2.73~eV, which is only 0.14~eV below the
figure obtained from EPR experiments,\cite{web80} and only 0.5~eV
lower than the enthalpy of formation of Fe$_{\mathrm{s}}$, suggesting
that the concentration of the latter defect could be relevant, particularly
in n-type Si, where it could act as a strong recombination center
for minority carriers in solar material.

Looking more closely at the Fe$_{\mathrm{i}}$ impurity, we confirm
that it gives rise to a single donor level, calculated at $E_{\mathrm{v}}+0.33$~eV
(only 0.05~eV below the well established transition measured by DLTS\cite{fei78,wun82}).
Inspection of further ionization and a comparison with the Fe$_{\mathrm{i}}$B$_{\mathrm{s}}$
defect allowed us to conclude that Fe$_{\mathrm{i}}(+/+\!+)$ is resonant
with the valence of the host. The calculated isomer shifts of neutral
and positively charged Fe$_{\mathrm{i}}$ in Si are calculated as
0.72~mm/s and 0.67~mm/s, respectively. We attribute the rather small
charge dependence of the IS to the nodal character of the 3d level
on Fe (which is the one involved in the donor transition), but also
to the fact that the Fe atom is in the 0 oxidation state, leaving
the 4s state (with amplitude on Fe) empty during the $(0/+)$ transition.
The agreement between the calculated IS for Fe$_{\mathrm{i}}^{+}$
and the available Mössbauer data in p-type Si is good. On the other
hand, for the neutral charge state, experiments came up with two rather
different values, $\delta=0.40$~mm/s or 0.77~mm/s, clearly suggestion
that further work is necessary in order to clarify the picture.

The iron-boron pair was also investigated. We started by looking at
the electronic structure and levels of the most stable form of this
defect, where the B atom replaces a Si first neighbor of the Fe$_{\mathrm{i}}$
defect (trigonal structure). In agreement with DLTS measurements we
obtained donor and acceptor levels at $E_{\mathrm{v}}+0.20$~eV and
$E_{\mathrm{c}}-0.35$~eV, respectively. The IS values for the $\mathrm{Fe_{i}B_{s}}$
pairs in charge state $q$ are found to be rather close to those of
Fe$_{\mathrm{i}}$ in the $q+1$ charge state. This supports the model
for the FeB pair as a Coulomb-stabilized complex composed by a Fe$_{\mathrm{i}}^{+}$
cation next to a B$_{\mathrm{s}}^{-}$ anion.

We confirm previous calculations\cite{est08} where Fe$_{\mathrm{s}}$
was predicted to be a deep acceptor. Our calculations indicate a Fe$_{\mathrm{s}}(-/0)$
level at 0.38~eV below $E_{\mathrm{c}}$. The calculated value for
the IS of Fe$_{\mathrm{s}}$ was $-0.13$~mm/s, in line with the
generally accepted value assigned to Fe$_{\mathrm{s}}$ ($-0.04$~mm/s)
in Mössbauer measurements. We also confirm that there is a strong
binding energy between Fe$_{\mathrm{i}}$ and a Si vacancy. The reaction
$\mathrm{Fe_{i}}+\mathrm{V}\rightarrow\mathrm{Fe_{s}}$ was found
to realease about 3~eV. We investigated this reaction in-depth using
the nudged elastic band method. In disagreement with the prevalent
view, we concluded that the reaction kinetics is effectively limited
by the migration rate of the vacancy, meaning that several spectroscopic
signals (including the NL19 center from EPR), which were detected
above room-temperature, were incorrectly connected to an iron-vacancy
pair.

Our final analysis concerned the iron-divacancy pair (Fe$_{\mathrm{i}}$V$_{2}$)
in Si, comprising a Fe atom at the center of a divacancy. The Fe$_{\mathrm{i}}$V$_{2}$
defect was predicted to be a very deep double acceptor with levels
at $\mathrm{Fe_{i}V_{2}}(-/0)=E_{\mathrm{v}}+0.10$~eV and $\mathrm{Fe_{i}V_{2}}(-/0)=E_{\mathrm{v}}+0.41$~eV.
We concluded that this is a rather stable complex and should be dominant
in the presence of vacancies, both after sample quenching from high
temperatures, and in samples that were subject to irradiation with
electrons and heavier particles. We anticipate that the Mössbauer
signal for the neutral complex is a doublet centered at $\delta=0.35$~mm/s
and split by $\Delta=1.1$~mm/s. In the negative charge state the
centroid increases slightly to 0.39~mm/s but the density becomes
more isotropic with $\Delta=0.4$~mm/s.

Finally, we studied the Mössbauer parameters for Fe$_{\mathrm{i}}$
and Fe$_{\mathrm{s}}$ in Ge, diamond, 3C-SiC and Si-rich SiGe alloys.
We confirm the observed approximate linear dependence of the IS with
the distance between Fe and its first neighbors. The agreement between
theory and experiments is very good in general. Additionally, we fill
in some \emph{blanks} and improve upon some previous calculations.
This includes the assignment of Fe$_{\mathrm{s}}$ and Fe$_{\mathrm{i}}$
defects with specific neighbors in SiC, as well as the IS for Fe$_{\mathrm{s}}$
in diamond.

\section*{Acknowledgements}

JC thanks Prof. Yutaka Yoshida for many fruitful discussions. This
work was funded by the Fundação para a Ciência e a Tecnologia (FCT)
under projects PTDC/CTM-ENE/1973/2012 and UID/CTM/50025/2013, and
funded by FEDER funds through the COMPETE 2020 Program. The authors
would like to acknowledge the contribution of the COST Action MP1406.
Computer resources were provided by the Swedish National Infrastructure
for Computing (SNIC) at PDC.

\bibliographystyle{apsrev4-1}


%

\end{document}